\documentclass[12pt]{article}

\usepackage{amsmath}
\usepackage{graphicx,psfrag,epsf}
\usepackage{enumerate}
\usepackage{natbib}
\usepackage{url} 
\setcitestyle{authoryear}
\usepackage[hidelinks]{hyperref}
\usepackage{xcolor}
\hypersetup{
    colorlinks,
    linkcolor={black!50!black},
    urlcolor={blue!80!black},
    citecolor = {black!80!black}
}
\usepackage{authblk}
\usepackage{bm}
\usepackage{dsfont}
\usepackage[ruled,vlined]{algorithm2e}
\usepackage{multirow}
\usepackage{multicol}
\usepackage{subcaption}
\usepackage{comment}
\usepackage{adjustbox}

\DeclareMathOperator*{\argmin}{arg\,min}
\usepackage{kantlipsum,setspace}
\usepackage[font=small]{caption}

\newcommand{\blind}{1}

\begin{document}

\def\spacingset#1{\renewcommand{\baselinestretch}%
{#1}\small\normalsize} \spacingset{1}

%%%%%%%%%%%%%%%%%%%%%%%%%%%%%%%%%%%%%%%%%%%%%%%%%%%%%%%%%%%%%%%%%%%%%%%%%%%%%%

\if1\blind
{
  \title{\bf{Bayesian Lesion Estimation with a Structured Spike-and-Slab Prior}}
  \author[1]{Anna Menacher}
  \author[2]{Thomas E. Nichols}
  \author[1]{Chris Holmes}
  \author[1,2]{Habib Ganjgahi\footnote{Please address any correspondence to \href{mailto:habib.ganjgahi@stats.ox.ac.uk}{habib.ganjgahi@stats.ox.ac.uk}.}}
  \affil[1]{\small{Department of Statistics, University of Oxford}}
  \affil[2]{\small{Nuffield Department of Population Health, University of Oxford}}
  \maketitle
} \fi

\if0\blind
{
  \bigskip
  \bigskip
  \bigskip
  \begin{center}
    {\LARGE\bf Bayesian Lesion Estimation with a Structured Spike-and-Slab Prior}
\end{center}
  \medskip
} \fi

\bigskip

\begin{abstract}
Neural demyelination and brain damage accumulated in white matter appear as hyperintense areas on T2-weighted MRI scans in the form of lesions. Modeling binary images at the population level, where each voxel represents the existence of a lesion, plays an important role in understanding aging and inflammatory diseases. We propose a scalable hierarchical Bayesian spatial model, called BLESS, capable of handling binary responses by placing continuous spike-and-slab mixture priors on spatially-varying parameters and enforcing spatial dependency on the parameter dictating the amount of sparsity within the probability of inclusion. The use of mean-field variational inference with dynamic posterior exploration, which is an annealing-like strategy that improves optimization, allows our method to scale to large sample sizes. Our method also accounts for underestimation of posterior variance due to variational inference by providing an approximate posterior sampling approach based on Bayesian bootstrap ideas and spike-and-slab priors with random shrinkage targets. Besides accurate uncertainty quantification, this approach is capable of producing novel cluster size based imaging statistics, such as credible intervals of cluster size, and measures of reliability of cluster occurrence. Lastly, we validate our results via simulation studies and an application to the UK Biobank, a large-scale lesion mapping study with a sample size of 40,000 subjects.

\end{abstract}

\noindent%
\vfill

\newpage
\spacingset{1.8} % DON'T change the spacing!
\section{Introduction}
\label{sec: introduction}
\label{sec: motivation}
\begin{figure}[!ht]
\centering
\begin{subfigure}{0.16\textwidth}
\includegraphics[width=\linewidth]{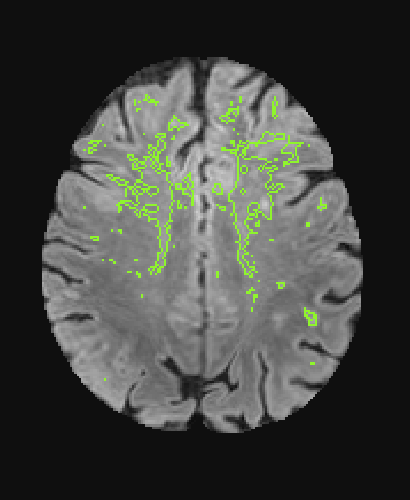}
\end{subfigure}
\begin{subfigure}{0.16\textwidth}
\includegraphics[width=\linewidth]{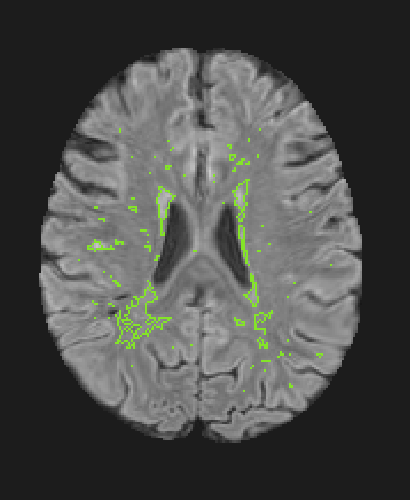}
\end{subfigure}
\begin{subfigure}{0.16\textwidth}
\includegraphics[width=\linewidth]{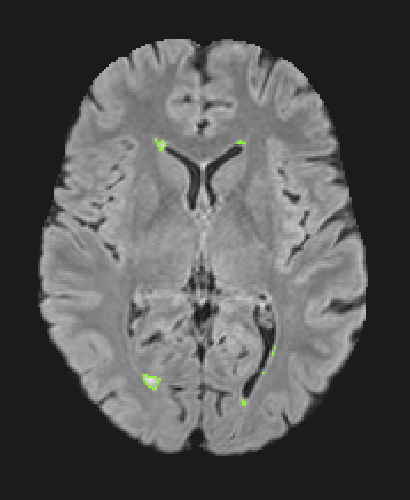}
\end{subfigure}
\begin{subfigure}{0.16\textwidth}
\includegraphics[width=\linewidth]{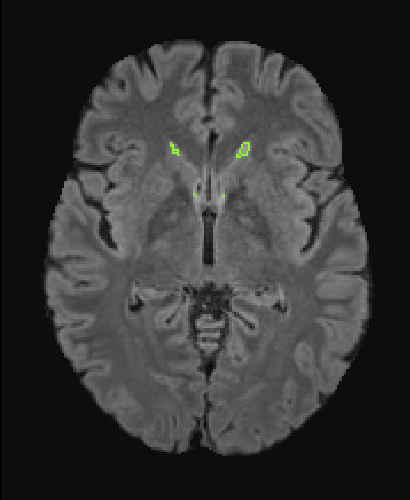}
\end{subfigure}

\caption{Contours of binary lesion masks from four healthy subjects from the UK Biobank with varying lesion numbers and sizes, where green outlines indicate lesions which show the heterogeneity of lesion incidence at various 2D axial slices from the 3D lesion mask, see Section 4.2 of the supplementary materials for further detail.}
\label{fig: lesion_mask}
\end{figure}

Magnetic resonance imaging (MRI) is a non-invasive imaging technique to study human brain structure and function. Accumulated damages to the white matter, known as lesions, appear as localized hypo-/ hyperintensities in MRI scans \citep{wardlaw2013}. The total burden of these lesions is often associated with cognitive disorders, aging and cerebral small vessel disease \citep{wardlaw2013, wardlaw2015}. Lesion prevalence is higher for older adults \citep{griffanti2018} and for individuals with cerebrovascular risk factors, such as hypertension, alcohol consumption or smoking history \citep{rostrup2012}. White matter lesions are also an overall indicator of poor brain health and have been found to triple the risk of stroke and double the risk of dementia and death and are associated with cognitive impairment, functional decline, sensory changes or motor abnormalities \citep{Debette2010}. Not all white matter lesions however are attributed to aging or an increased cerebrovascular risk burden. For example, white matter hyperintensities can also occur due to multiple sclerosis, Alzheimer's disease or as a result of a stroke \citep{Debette2010, prins2015}. While white matter lesions due to vascular origin are a result of chronically reduced blood flow and incomplete infarction leading to altered cerebral autoregulation, the non-vascular demyelination as seen in multiple sclerosis is caused by an autoimmune response against myelin proteins \citep{sharma2021}. Regardless of etiology, an important clinical feature is the spatial location of lesions; while noting that lesions exhibit a high level of variability, together with the size and number of lesions, for both between and within subjects, as seen in the binary lesion masks in Figure~\ref{fig: lesion_mask}. Elderly patients tend to present scattered lesions which later form to confluent lesions whereas white matter lesions of non-vascular origin have a particularly heterogeneous presentation where the disease course can result in rapid progression or alternation between relapses and remissions \citep{sharma2021}. Identifying spatial locations in the brain where lesion incidence is associated with different covariates (e.g. age, hypertension, cardiovascular disease) is known as lesion mapping and is an essential tool to locate the brain regions that are particularly vulnerable to damage from various risk factors and inform development of interventions to reduce incidence or severity of disease \citep{veldsman2020}.

\subsection{Mass-univariate Methods and Other Spatial Models}
\label{sec: other models}

The standard practice for lesion mapping is mass-univariate \citep{rostrup2012}. In this approach a logistic regression model is fitted at each voxel or spatial location independently, any form of spatial dependence among neighboring locations is ignored. Moreover, most methods fail to address the problem of complete separation which often occurs in logistic regression models when the output variable separates a subject-specific predictor variable or a combination of input features perfectly and hence leads to infinite and biased maximum-likelihood estimates \citep{firth1993}. This problem can be addressed with a logistic regression approach known as Firth Regression, which utilizes a penalized likelihood approach and produces mean-bias reduced parameter estimates \citep{firth1993, kosmidis2020}. 

Bayesian spatial models on the other hand are capable of accounting for the spatial dependence structure among neighboring voxels in a single joint model. For example, \cite{ge2014} have developed a Bayesian spatial generalized linear mixed model (\mbox{BSGLMM}) with a probit link function where the probability of lesion presence is modeled via a linear combination of fixed and random effects and subject-specific covariates. BSGLMM places a spatial smoothing prior directly on the parameters, specifically a conditional autoregressive model prior \citep{besag1974}, which may induce bias due to oversmoothing of regression coefficients. Moreover, BSGLMM relies on sequential Markov chain Monte Carlo (MCMC) methods for posterior computation which do not scale well to the large sample sizes found in the UK Biobank.

The last decade of brain imaging has brought immense insight into our understanding of the human brain. However, these findings suffer from small and unrepresentative samples. In addition, environmental and genetic factors that may explain individual differences are ignored which in turn undermines brain related findings. These limitations are being addressed in large-scale epidemiological studies, such as the UK Biobank or the ABCD study, by collecting data on thousands (instead of tens) of subjects. While the main advantage of these data sources lies in their larger sample sizes, they are also beneficial due to their inclusion of multiple high-dimensional imaging modalities as well as recording numerous environmental factors, neurocognitive scores, and other clinical data. The existing methods for brain mapping, specifically lesions, are either simplistic, ignoring complex spatial dependency, or are not scalable to large-scale studies.  

In order to address the limitations of previous methods, we propose a multivariate Bayesian model for lesion mapping in large-scale epidemiological studies that (1) uses variable selection and shrinkage priors, (2) takes into account the spatial dependency through a parameter that controls the level of sparsity rather than directly smoothing regression coefficients, and (3) relies on an approximate posterior sampling method based on Bayesian bootstrap techniques rather than MCMC, for parameter estimation and inference. Hence, this allows us to fit the model to thousands of subjects and appropriately account for the spatial dependency in lesion mapping studies containing over 50,000 voxel locations. We also want to acknowledge that other model choices in the literature may better capture the association between lesions and covariates \citep{li2021, zeng2022, whiteman2022}; however, we favor a model that enables us to scale parameter estimation and inference to large-scale epidemiological studies, see Section 13 of the supplementary material for a detailed literature review. 

\subsection{Bayesian Variable Selection}
\label{sec: spike-and-slab regression}
In this and the following subsection we will cover a short overview on the literature of Bayesian variable selection and approximate posterior sampling and refer readers to an in depth discussion in Section 14 of the supplementary materials. We utilize Bayesian variable selection to improve brain lesion mapping by shrinking small coefficients towards zero, thus helping with prediction, interpretation and reduction of spurious associations in high-dimensional settings. A commonly applied technique for Bayesian variable selection is spike-and-slab regression which aims to identify a selection of predictors within a regression model. The original spike-and-slab mixture prior places a mixture of a point mass at zero and a diffuse distribution on the coefficients \citep{Mitchell1988}. \cite{George1993, George1997} have increased the computational feasibility of spike-and-slab regressions by introducing a continuous mixture of Gaussians formulation where the spike distribution is defined by a normal distribution with a small variance rather than a point mass prior. The binary latent variable, sampled from a Bernoulli distribution with inclusion probability, determines which mixture component a variable belongs to and enables variable selection. Overall, the options of continuous shrinkage priors in the literature are large, see \cite{piironen2017} for a comparison of different methods.

The spike-and-slab regression is also able to incorporate spatial information, replacing the exchangeable Bernoulli prior on the inclusion indicator variables, with a structured spatial prior using a vector of inclusion probabilities. Previous examples of introducing structure within a spike-and-slab regression include the placement of a logistic regression product prior \citep{stingo2010} on the latents in order to group biological information for a genetics application, an Ising prior which incorporates structural information for a high-dimensional genomics application \citep{Li2010a} or a structured spike-and-slab prior with a spatial Gaussian process prior \citep{Andersen2014}.

\subsection{Approximate Posterior Inference and Sampling}
\label{sec: approximate posterior sampling}
The gold standard of parameter estimation and inference for spike-and-slab regression with a continuous mixture of Gaussians prior is Gibbs sampling \citep{George1993}. However, in high-dimensional regression settings as well as large sample size scenarios other more scalable approximate methods are required due to the intense computational burden. 

Expectation propagation (EP) \citep{Minka2001} or variational inference \citep{jordan1999} algorithms redefine the problem of approximating densities through optimization \citep{Blei2017}. Both of these methods have been extensively studied for spike-and-slab regression problems \citep{Hernandez-Lobato2013, Carbonetto2012}. The EP algorithm however poses several challenges as it is computationally intensive for even moderate sample sizes, there is no guarantee of convergence, and its poor performance for multimodal posteriors due to the problematic need to incorporate all modes in its approximation \citep{Bishop2006}. Poor variational approximations can arise due to slow convergence, a simplistic choice of variational families, or due to underestimation of the posterior variance as the KL-divergence tends to under-penalize thin tails \citep{yao18a}. 

In neuroimaging applications, however, we require accurate uncertainty estimates and hence we use approximate posterior sampling which captures the marginal posterior density more accurately than variational densities while remaining to be highly scalable due to embarrassingly parallel implementations \citep{fong2019}. The cornerstone of these methods lies in the Bayesian bootstrap \citep{rubin1981} and the Weighted Likelihood Bootstrap (WLB) \citep{newton1994}. The WLB randomly re-weights the likelihood with Dirichlet weights for the observations and maximizes this likelihood with respect to the parameter of interest. Using WLB, \cite{lyddon2018} and \cite{fong2019} developed Bayesian nonparametric learning (BNL) routines which utilize parametric models to achieve posterior sampling through the optimization of randomized objective functions. 

Our focus lies on the recently introduced method by \cite{nie2021bayesian} which combines Bayesian bootstrap methods with a new class of jittered spike-and-slab LASSO priors and obtains samples via optimization of many independently perturbed datasets by re-weighting the likelihood and by jittering the prior with a random mean shift. This procedure is equivalent to adding pseudo-samples from a prior sampling distribution as in the case of BNL \citep{fong2019}. We argue that for high-dimensional datasets with large samples, where memory allocation is already a computational concern, the approach by \cite{nie2021bayesian} is favorable as it merely requires storing a set of mean shift parameters compared to an arbitrarily large number of pseudo-samples \citep{fong2019}. 

The remainder of this article is organized as follows. In Section 2, we formulate a Bayesian spatial spike-and-slab regression model with approximate posterior sampling via a Bayesian bootstrap procedure. We then assess the quality of our method, called BLESS, via simulation studies in Section~3 and give the results from the UK Biobank application in Section~4. We conclude the paper with a discussion in Section 5 and provide further details on the variational distributions as well as more simulation study results in the supplementary material. Additionally, we make all code publicly available on Github \footnote{\url{https://github.com/annamenacher/BLESS}}.
\section{Methods}
\label{sec: model}

Our model for \textbf{B}ayesian \textbf{L}esion \textbf{E}stimation with a \textbf{S}tructured \textbf{S}pike-and-Slab (\textbf{BLESS}) prior is formulated as a Bayesian spatial hierarchical generalized linear model. While we specifically focus on neuroimaging applications within this paper, the model can be applied to any form of spatial binary data on a lattice and can equally be extended to various neuroimaging modalities other than lesion masks. 

Throughout this paper we use boldface to indicate a vector or matrix. We model the binary data $y_i(s_j)$ for every subject $i = 1, \dots, N$ at voxel location $s_j \in \mathcal{B} \in \mathds{R}^3$, $j = 1, \dots, M$, with a Bernoulli random variable with lesion probability $p_i(s_j)$. Due to computational reasons, we choose to model the binary data via a probit link function which defines the relationship between the conditional expectation $\eta_i(s_j)$ and the linear combination of input features $\bm{x}_i$ containing $P$ subject-specific covariates, spatially-varying parameters $\bm{\beta}(s_j)$ and a spatially-varying intercept $\beta_0(s_j)$. While the data comprise an image for each subject, we store the data as unraveled $M$-vectors $\bm{y}_i$ for each subject or $N$-vectors $\bm{y}(s_j)$ for each voxel. 

The Bayesian spatial generalized linear model for subject $i$ at location $s_j$ is specified as 
\begin{align}
\label{eq: 1}
    [y_i(s_j) | p_i(s_j)] &\sim \text{Bernoulli}[p_i(s_j)] \\
	\Phi^{-1}\left(  p_i(s_j) \right) &= \eta_i(s_j) = \bm{x}_i^{\text{T}} \bm{\beta}(s_j) + \beta_0(s_j),
 \label{eq: 2}
\end{align} 
where the equations reflect the random and systematic component, respectively and the link function is given by the cumulative Gaussian density $\bm{\Phi}(\cdot)$.  

Furthermore, we reparameterize the Bayesian probit regression model defined in Equation (\ref{eq: 1}) and (\ref{eq: 2}) exactly via the data augmentation approach by \cite{albert1993} by introducing latent normal variables in Equation (\ref{eq: 3}) and (\ref{eq: 4}) into the model in order to ease the computational complexity. This approach assumes that the probit regression has an underlying normal regression structure on latent continuous data. These independent continuous latent variables $z_i(s_j)$ for every subject $i = 1, \dots, N$ and $j = 1, \dots, M$ are drawn from the following normal distribution 
\begin{equation}
\label{eq: 3}
    z_i(s_j) | \eta_i(s_j) \sim \mathcal{N}(\eta_i(s_j), 1)
\end{equation}

 where the conditional probability of $y_i(s_j) = 1$ is given by 
\begin{align}
\label{eq: 4}
	\Pr[y_i(s_j) = 1 | z_i(s_j)] &=
\begin{cases}
1, & z_i(s_j) > 0,\\
0, & z_i(s_j) \leq 0.
\end{cases}
\end{align}
\subsection{Prior Specifications}
\label{sec: priors}

We build a Bayesian hierarchical regression model by placing a continuous version of a spike-and-slab prior on the spatially-varying $P$-coefficient vector $\bm{\beta}(s_j)$. The continuous mixture of Gaussians with two different variances, consisting of the spike and the slab distribution, is given by
\begin{equation}
 \beta_p(s_j)\ |\ \gamma_p(s_j)\ \sim\ \mathcal{N}(0,\ \nu_1 \gamma_p(s_j)\ +\ \nu_0 \left[1-\gamma_p(s_j)\right]), \end{equation}
 where $\gamma_p(s_j)$ is a latent binary indicator variable for covariate $p = 1, \dots, P$ and locations $j = 1, \dots, M$, $\nu_0$ is the spike variance and $\nu_1$ is the slab variance which determine the amount of regularization. Variable selection is implemented via the latents $\gamma_p(s_j)$, and specifically it localizes the spatial effect of each variable. Due to the continuous spike-and-slab specification, the variance within the spike distribution is always $\nu_0 > 0$ which ensures the continuity of the spike distribution and therefore the derivation of closed form solutions of the variational parameter updates. The slab variance on the other hand is a set to a fixed value to include the range of all possible values of the spatially-varying coefficients. The combination of a small spike variance and a large slab variance with latent indicator variables for every covariate and location introduces a selective spatial shrinkage property that shrinks smaller coefficients close to zero and leaves the large parameters unaffected. 
 
In order to account for the spatial dependence across the brain, we place an independent logistic regression prior with non-exchangeable inclusion probabilities on the latent binary indicator variables $\bm{\gamma}(s_j)$ sampled from a Bernoulli distribution, similar to \cite{stingo2010}. The prior is non-exchangeable because we incorporate structural information via the sparsity parameter $\bm{\theta}(s_j) \in \mathds{R}^P$ which ensures that certain voxel locations are more likely to be included in the model than others. In the context of brain imaging data this means that voxels that are nearby each other probably have a similar inclusion probability. Specifically, we model the latents $\gamma_p(s_j)$ via
 \begin{equation}
  \gamma_p(s_j)\ |\ \theta_p(s_j)\ \sim\ \text{Bernoulli}(\sigma\left[\theta_p(s_j)\right]) ,
  \end{equation}
  where $\sigma(\cdot)$ is a sigmoid function.
  
 The hierarchical spatial regression model is completed by placing a spatial prior on the sparsity parameter $\bm{\theta}^{\text{T}} = \left[\bm{\theta}^{\text{T}}(s_1), \dots, \bm{\theta}^{\text{T}}(s_M)\right]$: a length $PM$ column vector. We choose a multivariate conditional autoregressive (MCAR) prior as a spatial prior due to computational reasons \citep{gelfand2003b}. Alternative priors could be considered in lieu of this type of a simple smoothing prior; however, this would significantly increase the computational complexity of the model at hand.  

The full conditional distribution for $\bm{\theta}(s_j)$ is given by the following multivariate normal distribution and utilizes the notation defined by \cite{mardia1988}:
 
 \begin{equation}
 \left[\bm{\theta}(s_j)\ |\ \bm{\theta}(-s_j), \bm{\Sigma}^{-1} \right]\ \sim\ \text{MVN}\left(\frac{\sum_{s_r \in \partial s_j} \bm{\theta}(s_r)}{n(s_j)},\ \frac{\bm{\Sigma}}{n(s_j)}\right),\\
 \end{equation}
 where $\bm{\Sigma}$ is a symmetric positive definite smoothing matrix. The sum $\sum_{s_r \in \partial s_j}$ defines the sum over the neighborhood voxels at location $s_j$, $\partial s_j$ defines the set of neighbors at location $s_j$ and $n(s_j)$ is the cardinality of the neighborhood set. For our MRI scans we consider only neighbors sharing a face, so therefore most of the interior of the brain has $n(s_j)=6$ neighbors whereas locations near the brain mask have $n(s_j) < 6$. 
 
We then describe the joint distribution over the sparsity parameters, up to a proportionality constant, by utilizing Brooks's lemma \citep{brook1964} which is given by:
\begin{equation}
  \pi(\bm{\theta}\ |\ \bm{\Sigma})\ \propto\ \exp\left\{-\frac{1}{2}\sum_{s_j \sim s_{j'}} [\bm{\theta}(s_{j}) - \bm{\theta}(s_{j'})]^{\text{T}}\bm{\Sigma}^{-1}[\bm{\theta}(s_j) - \bm{\theta}(s_{j'})]\right\} ,
\end{equation}
where the sum $\sum_{s_j \sim s_{j'}}$ describes the sum over neighborhood voxels, and $s_j \sim s_{j'}$ indicates that $s_j$ and $s_{j'}$ are neighbors. This joint prior distribution is improper and not identifiable according to \cite{besag1986}. However, the posterior of $\bm{\theta}$ is proper, if there is information in the data with respect to the sparsity parameters. Lastly, we finish specifying the Bayesian hierarchical regression model by placing an uninformative, conjugate Wishart prior over the precision matrix $\bm{\Sigma}^{-1}$ to fully specify the model with
\begin{equation}
\bm{\Sigma}^{-1} \sim \text{Wishart}(\nu,\ \bm{I}),
\end{equation}
where the degrees of freedom are given by $\nu = P$ and the scale matrix is defined by the identity matrix $\bm{I}$ \citep{ge2014}. 

\subsection{Posterior Approximation}
\label{sec: VI}
The first element of our scalable approximate posterior sampling approach is a variational approximation to the posterior using optimization instead of MCMC sampling. Every sample within the approach in Section \ref{sec: BB-BLESS} is acquired by optimizing the posterior via variational inference. We opt for variational inference due to the non-conjugacy in the hierarchical model induced by specifying a logistic function around the sparsity parameters $\bm{\theta}$ in the inclusion probabilities of the spike-and-slab priors. Local variational approximations solve this problem by finding a bound on an individual set of variables via a first-order Taylor approximation \citep{Jaakkola2000}. For general variational inference, we then require the full joint distribution of the Bayesian spatial regression model, consisting of the likelihood $p(\bm{Y} | \bm{X}, \bm{\beta}, \bm{\beta_0})$ and the joint prior $p(\bm{Z},\bm{\beta},\bm{\beta_0}, \bm{\gamma}, \bm{\theta}, \bm{\Sigma}^{-1})$, which is given by
\begin{align}
p(\bm{Y}, \bm{Z}, \bm{X}, \bm{\beta}, \bm{\beta}_0, \bm{\gamma}, \bm{\theta}, \bm{\Sigma}^{-1}) =&\ p(\bm{Y} | \bm{Z}) p(\bm{z} | \bm{X}, \bm{\beta}, \bm{\beta}_0) p(\bm{\beta}_0) p(\bm{\beta} | \bm{\gamma}) p(\bm{\gamma} | \bm{\theta})\\
&\ p(\bm{\theta} | \bm{\Sigma}^{-1}) p(\bm{\Sigma}^{-1}) \nonumber.
\end{align}

We write the entire set of model parameters as $\bm{\Psi} = \left\{\bm{Z}, \bm{\beta}, \bm{\beta_0}, \bm{\gamma}, \bm{\theta}, \bm{\Sigma}^{-1}\right\}$ where the conditional distribution of each model parameter $\bm{\psi}$ is obtained as $p(\bm{\psi} | \bm{y}) = \frac{p(\bm{\psi}, \bm{y})}{p(\bm{y})}$. We acquire an approximation to the exact posterior by firstly specifying a family of densities~$\mathcal{Q}$ over each model parameter $\psi_j$ and secondly identifying the parameters of the candidate distribution $q(\psi_j) \in \mathcal{Q}$ that minimizes the Kullback-Leibler (KL) divergence, given by
\begin{equation}
    q^*(\psi_j) = \argmin_{q(\psi_j) \in \mathcal{Q}} \text{KL}\ \{q(\psi_j)\  ||\  p(\psi_j | \bm{y})\}.
\end{equation}
We aim to minimize the difference between the exact posterior $p(\psi_j | \bm{y})$ and the variational distribution $q(\psi_j)$ to find the best approximate distribution $q^*(\psi_j)$. However, rather than computing the KL-divergence which contains the log-marginal of the data, a quantity that is often not computable, we optimize the evidence lower bound (ELBO) \citep{Blei2017}
\begin{equation}
    \mathcal{L}(q) 
    \geq \mathds{E}_{q(\bm{\Psi})}\left[\ln\left\{p(\bm{Y},\bm{X},\bm{\Psi})\right\}\right] - \mathds{E}_{q(\bm{\Psi})}\left[\ln\left\{q(\bm{\Psi})\right\}\right] . 
\end{equation}
The derivation of the variational distributions and the ELBO can be found in supplementary material in Section 2.2 and 2.3 respectively. The variational density $q_j(\psi_j)$ is derived by taking the exponentiated expected log of the complete conditional given all the other parameters and the data which is defined by $q_j(\psi_j) \propto \exp\{\mathds{E}_{-j}[\log\{p(\psi_j |\bm{\psi}_{-j}, \bm{X})\}]\}$ where the expectation is over the fixed variational density of other variables $\bm{\psi}_{-j}$, given by $\prod_{\ell \neq j} q_{\ell}(\psi_{\ell})$. By determining the variational distributions~$q$, we successively update each parameter $\bm{\psi}$, while holding the others fixed, via mean-field coordinate ascent variational inference \citep{Bishop2006}. Further details on initialization and convergence of variational inference can be found in the supplements. 

\subsection{Dynamic Posterior Exploration}
\label{sec: DPE}
Dynamic posterior exploration (DPE) \citep{Rockova2014}, is an annealing-like strategy, which fixes the slab variance to a large, fixed value. The procedure works by starting in a smooth posterior landscape and aims to discover a sparse, multimodal posterior by gradually decreasing the value of the spike parameter until it approximates the spike-and-slab point mass prior. When the starting spike variance is large, we should be able to easily identify a small set of local optima by maximizing the ELBO. Thereafter, the technique uses the result as a warm start for the next optimization with a reduced spike variance which leads to a more peaked posterior until the last value within a range of spike variances is evaluated and a stable solution to the optimization problem is found.

The process of dynamic posterior exploration can be split into three parts. Firstly, we perform parameter estimation via variational inference over a sequence of $K$ increasing spike variances $\nu_0 \in V = \{\nu_{0}^{(1)}, \dots, \nu_{0}^{(K)}\}$. After the initial evaluation of the backwards DPE procedure with $\nu_{0}^{(K)} \leq \nu_1$, every subsequent optimization is run with a successively smaller $\nu_0$ and initialized with the previously estimated variational parameters as a ``warm start'' solution. Secondly, the output of every optimization run within the sequence of spike parameter values $V$ is thresholded via the posterior inclusion probabilities. The thresholding rule for BLESS is based on the following inclusion probabilities
\begin{align}
\hat{\gamma}_p(s_j) &=  \begin{cases}
1, &  \text{if}\ P(\gamma_p(s_j) = 1 | \bm{Y}, \hat{\bm{\beta}}, \bm{\hat{\beta}_0}, \bm{\hat{\theta}}) > 0.5,\\
0, &  \text{if}\ P(\gamma_p(s_j) = 1 | \bm{Y}, \hat{\bm{\beta}}, \bm{\hat{\beta}_0}, \bm{\hat{\theta}}) \leq 0.5,
\end{cases} 
\label{eq: thresholding_rule}
\end{align}
which is equivalent to the local version of the median probability model defined by \cite{Barbieri2021}. Furthermore, the determination of active versus inactive voxels based on the inclusion probability $P(\gamma_p(s_j) = 1 | \bm{y})$ is equivalent to thresholding the parameter estimates~$\hat{\bm{\beta}}$ themselves where the threshold is given by the intersection of the weighted mixture of the spike-and-slab priors \citep{George1993, Rockova2014}. For BLESS, we choose the former thresholding rule based on the posterior inclusion probabilities considering that thresholding the parameters~$\hat{\bm{\beta}}$ would require the calculation of a different set of intersection points for every coefficient due to the non-exchangeable nature of the spatial prior within the inclusion probability. Thirdly, the estimated posterior with the smallest spike variance $\nu_0$ within the range of parameters $V$ is used. We do not assert that this $\nu_0$ is optimal per se, but that our annealing-like strategy obviates the need for a precise determination of $\nu_0$ as the estimates for the larger effects tend to stabilize at a particular solution of variational posterior parameters.

\begin{figure}[!ht]
\centering
\begin{subfigure}{0.35\textwidth}
\caption{Regularization Plot}
\includegraphics[width=\linewidth]{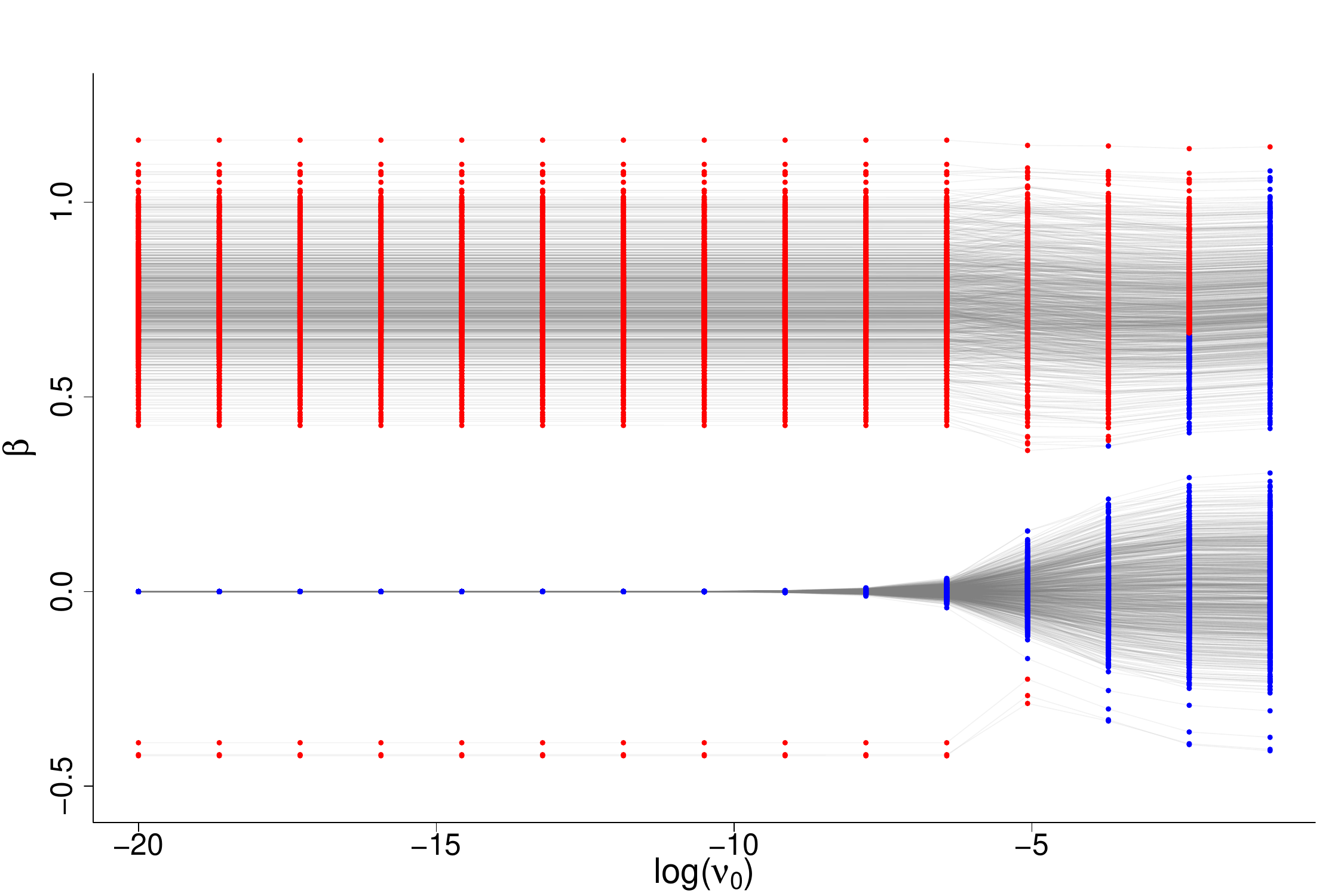}
\end{subfigure}
\begin{subfigure}{0.35\textwidth}
\caption{Marginal Plot}
\includegraphics[width=\linewidth]{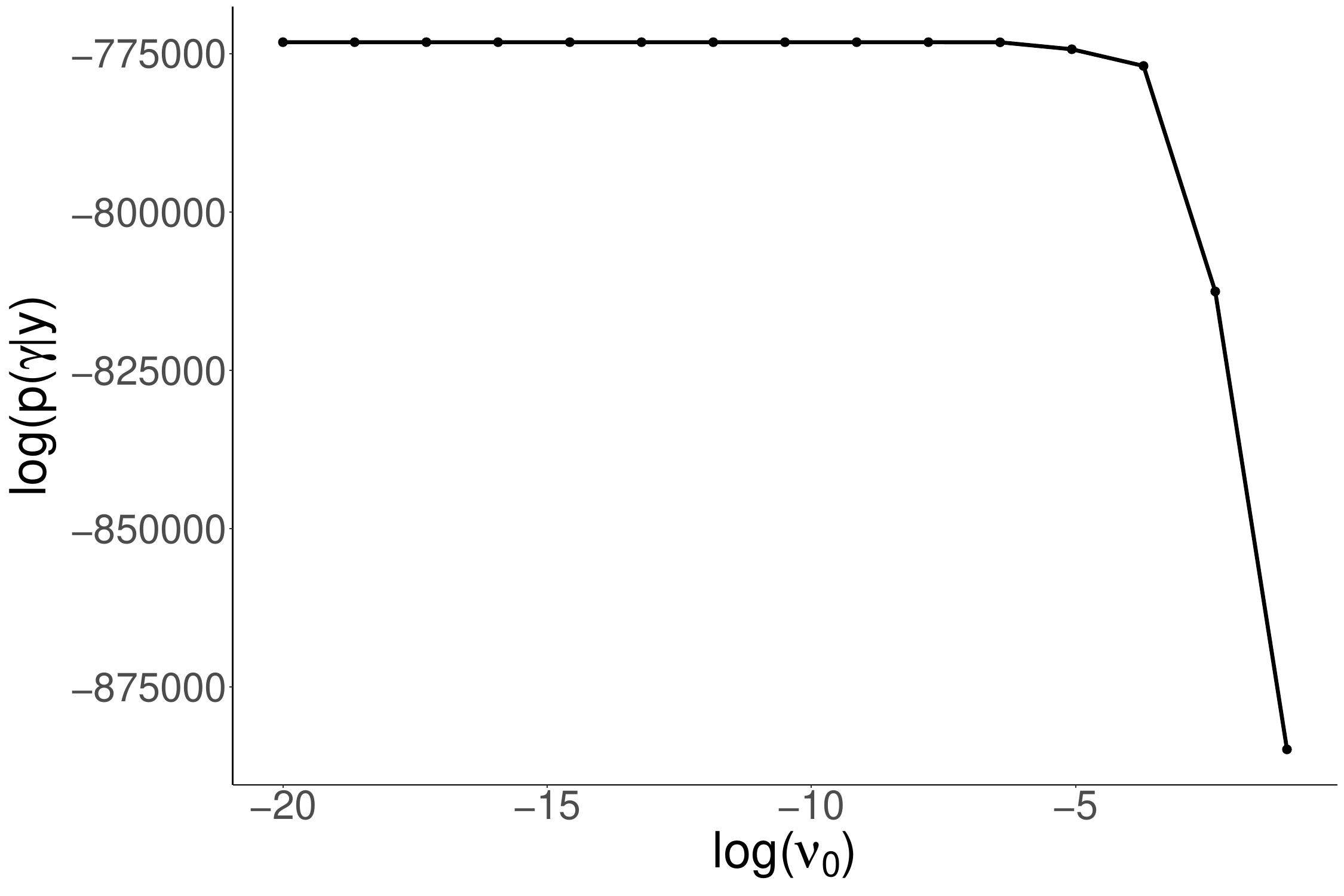} 
\end{subfigure}
\caption{(a) Regularization plot (active voxel: red, inactive voxel: blue) and (b) plot of marginal posterior of $\bm{\hat{\gamma}}$ under $\nu_0 = 0$ over a sequence of equidistant $\nu_0 \in V$ within log-space for the simulation study described in Section \ref{sec: simulation study} for sample size $N=$\ 1,000 and base rate intensity $\lambda=3$. Both plots indicate that parameter estimates have stabilized past spike variances of $\log(\nu_0) \leq -6$ within the DPE procedure. The regularization plot also shows how negligible (blue) coefficients are progressively shrunk towards 0 while the larger (red) coefficients remain almost unregularized.}
\label{fig: reg_plots}
\end{figure}

This behavior can be validated by two types of plots. Regularization plots enable the examination of the estimated coefficients over a sequence of spike variances. For each $\nu_0$, the color of the parameter values indicates whether or not a variable is included in (red) or excluded from (blue) the model based on the thresholded posterior probability of inclusion. Figure \ref{fig: reg_plots}(a) illustrates how the negligible coefficients are drawn to zero as the values of $\nu_0$ decrease, while the larger parameters of the active voxels stabilize and are unaffected by regularization. Hence, for the plot in Figure \ref{fig: reg_plots}(a) this occurs at a log-spike variance $\log(\nu_0) \leq -6$ where a local optimum has been identified and any further decrease in spike variance only leads to further shrinkage of the negligible coefficients. 

A complementary plot, especially useful when overplotting makes the regularization plot difficult to interpret, is the log-marginal posterior plot $\ln\{\pi_{\nu_0=0}(\bm{\gamma} | \bm{y}, \bm{\theta}, \bm{\Sigma}^{-1})\}$ for the latents. The maximum value of this quantity yields the posterior closest to approximating the point mass prior which is the goal of backwards DPE. Since our model contains intractable integrals, we use a variational approximation to the marginal posterior of $\bm{\gamma}$ under the prior of $\nu_0 = 0$. We utilize Jensen's inequality to bound the marginal probability integrating out the parameters $\bm{\beta}$, $\bm{\beta_0}$ and the latent variables $\bm{Z}$ via their respective variational approximation. The other model parameters $\bm{\theta}$ and $\bm{\Sigma}^{-1}$ are regarded as nuisance parameters. Specifically, the log-marginal posterior under $\nu_0 = 0$ and its approximation
\begin{align}
\ &\ln\{\pi_{\nu_0 = 0}(\bm{\gamma} | \bm{Y}, \bm{\theta}, \bm{\Sigma}^{-1})\} \\
=\ & \ln\left\{\int \int \int q(\bm{Z}, \bm{\beta}_{\gamma}, \bm{\beta_0}) \frac{p(\bm{Y},\bm{Z},\bm{\beta}_{\gamma},\bm{\beta_0},\bm{\gamma},\bm{\theta},\bm{\Sigma}^{-1}|\bm{X}_{\gamma})}{q(\bm{Z},\bm{\beta}_{\gamma},\bm{\beta_0})}d\bm{Z} d\bm{\beta}_{\gamma} d\bm{\beta_0}\right\} 
\end{align}
\begin{align}
\geq\ & \mathds{E}_{q(\bm{Z},\bm{\beta}_{\gamma},\bm{\beta_0})}\left[ \ln\left\{ p(\bm{Y},\bm{Z},\bm{\beta}_{\gamma},\bm{\beta_0},\bm{\gamma},\bm{\theta},\bm{\Sigma}^{-1}|\bm{X}_{\gamma})\right\}\right] -\\
\ &\mathds{E}_{q(\bm{Z},\bm{\beta}_{\gamma},\bm{\beta_0})}\left[ \ln\left\{ q(\bm{Z},\bm{\beta}_{\gamma},\bm{\beta_0})\right\}\right], \nonumber
\end{align}
where $\bm{\beta}_{\gamma} = \bm{\beta}\bm{\gamma}$, can be used to determine whether or not the parameters identified as active have stabilized by checking a single quantity rather than the solution path of all parameters of the model. Figure \ref{fig: reg_plots}(b) illustrates the marginal $\ln\{\pi_{\nu_0=0}(\bm{\gamma} | \bm{y}, \bm{\theta}, \bm{\Sigma}^{-1})\}$, showing a plateau for any log-spike variance $\log(\nu_0) \leq -6$ indicating a good approximation of the point mass prior. The marginal plot can be used in an equivalent manner to the regularization plot as a sanity check for visualizing the stabilization of large effects and continued shrinkage for the negligible coefficients at the end of the annealing-like process. 

\subsection{Uncertainty Quantification via Bayesian Bootstrap - BLESS}
\label{sec: BB-BLESS}

 \begin{algorithm}[H]
\KwResult{Sample of parameter estimates $\bm{\Tilde{\beta}}$ from approximate posterior distribution.}
 \textbf{Set:} $\nu_0 = \nu_0^{DPE}$; $\nu_1$: large, fixed value; $\alpha$: concentration parameter;\\ \ \ \ \ \ \ \ $\epsilon$: convergence criterion; $B$: number of bootstraps\\
 \For{b = 1, \dots, B}{
 1) Sample weights $\bm{w}^{(b)} \sim N \times Dirichlet(\alpha, \dots, \alpha)$.\\
 2) Sample mean shifts $\bm{\mu}^{(b)}(s_j)$ for all $j=1, \dots, M$ from $\mu_p(s_j) \sim \mathcal{N}(0, \nu_0)$.\\
 3) Calculate $\bm{\Tilde{\beta}}^{(b)}$ by acquiring variational posterior mean via approximating pseudo-posterior via variational inference.  
 }\caption{BB-BLESS}
 \label{alg: bb-bless}
\end{algorithm}
\bigskip
We use Bayesian bootstrapping techniques \citep{nie2021bayesian} to obtain approximate posterior samples. The approach for Bayesian Bootstrap - BLESS (BB-BLESS) is threefold and is described by Algorithm~\ref{alg: bb-bless}: Firstly, weights $w_i^{(b)}$ are sampled for every observation $i = 1, \dots, N$ from a Dirichlet distribution, which has been scaled by the sample size $N$, in order to re-weight the likelihood following the weighted likelihood bootstrap by \cite{newton1994}. Every weight $w_i^{(b)}$ hereby perturbs the contribution of each observation to the likelihood. Secondly, a prior mean shift $\mu_p(s_j)$ is sampled for every covariate $p =1, \dots, P$ and voxel $j = 1, \dots, M$  from the spike distribution. For initialization, the variational parameters estimated when performing DPE for BLESS-VI (BLESS estimated via variational inference alone) at the target spike variance value $\nu_0$ are used as initial values to BB-BLESS. This prior mean shift $\mu_p(s_j)$ is then used to center the prior for $\pi(\bm{\beta}(s_j) | \bm{\gamma}(s_j))$ on $\bm{\mu}(s_j)$ instead of $\bm{0}$ \citep{nie2021bayesian}. This combination of Bayesian bootstrap methods and the jittering of the spike-and-slab prior allows for approximate posterior sampling by repeatedly optimizing the updated ELBO with respect to its variational parameters to approximate a posterior density. The variational posteriors for all other nuisance parameters are also re-fitted for every bootstrap sample. Thirdly, we acquire a sample $\Tilde{\bm{\beta}}^{(b)}(s_j)$ by optimizing the ELBO with respect to the spatially-varying coefficient~$\Tilde{\bm{\beta}}^{(b)}(s_j)$. The following is a variational approximation to the pseudo-posterior, defined by a re-weighted likelihood and perturbed prior:
\begin{align}
\label{eq: bb_bless}
    q(\Tilde{\bm{\beta}}^{(b)}(s_j)) &\propto \exp\left\{\mathds{E}_{q(\bm{Z}, \bm{\beta_0}, \bm{\gamma}, \bm{\theta}, \bm{\Sigma}^{-1})}\left[ \ln\left\{ \prod_{i=1}^N p(y_i(s_j) | z_i(s_j))\ p(z_i(s_j) | \bm{\beta}(s_j), \bm{\beta_0}, \bm{x}_i)^{w_i^{(b)}} \right. \right. \right.\\
&\left. \left. \left. p(\bm{\beta}(s_j) | \bm{\mu}(s_j)^{(b)}, \bm{\gamma}(s_j))\  p(\bm{\beta_0})\ p(\bm{\gamma} | \bm{\theta})\ p(\bm{\theta} | \bm{\Sigma}^{-1})\ p(\bm{\Sigma}^{-1})  \vphantom{\prod_1^N} \right\} \right] \right\}, \nonumber 
\end{align}
where the Dirichlet weights are $(w_1^{(b)}, \dots, w_N^{(b)}) \sim \text{Dir}(\alpha, \dots, \alpha)$ and the jitter is drawn via the spike distribution $\mu_p(s_j) \sim \mathcal{N}(0, \nu_0)$. Each bootstrap sample $\Tilde{\bm{\beta}}^{(b)}(s_j)$ is acquired by taking the marginal variational posterior mean of the pseudo-posterior defined in Equation~(\ref{eq: bb_bless}) where the nuisance parameters are approximately marginalized out. Note that we prefer the variational posterior mean opposed to the maximum-a-posteriori (MAP) estimate for each bootstrap draw due to the computational tractability of the former. Using the MAP estimate would result in having to use numerical optimization at each iteration as some updates do not have a closed-form solution. We also acknowledge that while we do not provide theoretical guarantees within this paper, we do validate our work with numerical simulations. The full derivations of this method can be found in Section 3 of the supplementary material.
\section{Simulation Study}
\label{sec: simulation study}
In this section, we firstly explain the process of simulating lesion data where the ground truth is known. We perform various simulation studies to assess the performance of BLESS, estimated via variational inference (BLESS-VI), approximate posterior sampling (BB-BLESS) and traditional Gibbs sampling (BLESS-Gibbs), by assessing their marginal posterior distributions and quantities. In addition, we compare parameter estimates and predictive performance of our method to the mass-univariate approach, Firth regression \citep{firth1993}, and the Bayesian spatial model, BSGLMM \citep{ge2014}. For comparison, the latter is adopted to fit a Bayesian hierarchical modeling framework, similar to BLESS, where we add a spatially-varying intercept $\bm{\beta}_0(s_j)$ to match the setup of BLESS. For Firth regression, which fits an independent probit regression model with a mean bias reduction for every voxel location, we use the \textit{R} package \textit{brglm2} \citep{kosmidis2021}.

The main aim of many neuroimaging studies lies in the provision of accurate inference results. We therefore tailor the assessment of simulation studies on the evaluation of inference results rather than on coverage probabilities. We compare inference results by assessing true positive (TP), false positive (FP), true negative (TN), and false negative (FN) discoveries in the following measures: (1) sensitivity/true positive rate (TPR = $\frac{\text{TP}}{\text{TP} + \text{FN}}$), (2) true discovery rate (TDR = $\frac{\text{TP}}{\text{TP} + \text{FP}}$), (3) specificity/1 - false positive rate (FPR = $\frac{\text{FP}}{\text{FP} + \text{TN}}$), and (4) false discovery rate (FDR = $\frac{\text{FP}}{\text{FP} + \text{TP}}$). Lastly, we provide extensive simulation studies on the performance of BLESS-VI compared to a frequentist, mass-univariate approach as well as a Bayesian spatial model with a simulation study addressing varying sample sizes~$N$, base rate intensities~$\lambda$, and sizes of effect within an image. Base rate intensities hereby provide an indicator for the magnitude of various regression coefficient effect sizes where a smaller~$\lambda$ value yields smaller regression coefficients.  

For simulating the data, we adopt a data generating process that is different from our model in order to guarantee a fair comparison between the method we propose, BLESS, to the other methods, BSGLMM and Firth regression. We therefore use a data generating mechanism which simulates homogeneous regions of lesions proposed by \cite{ge2014}, with intensities that vary over subjects, which provides us with a tool to provide a fair comparison among the three methods evaluated.  
For our study, we consider $P=2$ effects in addition to an intercept, we label sex and group (e.g.~patient and control). We simulate 2-D binary lesion masks of size $50 \times 50$, $M=2,500$, with homogeneous effects in each $25 \times 25$ quadrant. The effect of sex leads to 4 times more lesions on the right side of an image for female subjects compared to the baseline. The second effect of group membership introduces an effect of 4 times more lesions within the lower left quadrant of an image for subjects within group 2. A Poisson random variable with base rate $\lambda$ determines the number of lesions. Further details and plots of the simulated data can be found in the supplementary materials in Section 5 alongside further sensitivity analyses, such as extending the neighborhood definition of the MCAR prior, in Section 10. We also use a simulation framework developed by \cite{KINDALOVA2021118090} to generate realistic looking lesion masks utilizing summary statistics from Firth regression based on UK Biobank data as truth. The results are similar and can be found in the supplements in Section~12.
\subsection{BB-BLESS Simulation Study}
\label{sec: bb_bless_sim_study}
In this simulation study on a low base rate and sample size scenario ($N=500$, $\lambda=1$) as well as a high base rate and sample size scenario ($N=$\ 1,000, $\lambda=3$), we want to assess the performance of BLESS-VI, BB-BLESS and BLESS-Gibbs on two scenarios with small and large regression coefficients based on their base rate intensity $\lambda=1$ and $\lambda=3$. The posterior quantities of BLESS-VI are acquired by running a separate backwards dynamic posterior exploration procedure for every dataset with an equispaced spike sequence of $\nu_0 = \exp\{-20, \dots, -1\}$ of length 15 and a slab variance of $\nu_1 = 10$. The method is initialized with the coefficients of Firth regression where we use the parameter estimates and respective inference results from the final run in the backwards DPE procedure ($\nu_0 = \exp(-20)$). We estimate BB-BLESS by drawing $B=1,000$ bootstrap replicates and Dirichlet weights with a concentration parameter $\alpha=1$. We run the Gibbs sampler for 15,000 iterations and discard 5,000 iterations as burn-in. The performance of BB-BLESS and BLESS-Gibbs is then greatly improved by utilizing the output of the backwards DPE procedure as parameter initialization for the respective parameter estimation techniques. 

\begin{figure}[!ht]
\centering
\begin{subfigure}{0.24\textwidth}
\caption{Inactive Voxel}
\includegraphics[width=1\linewidth]{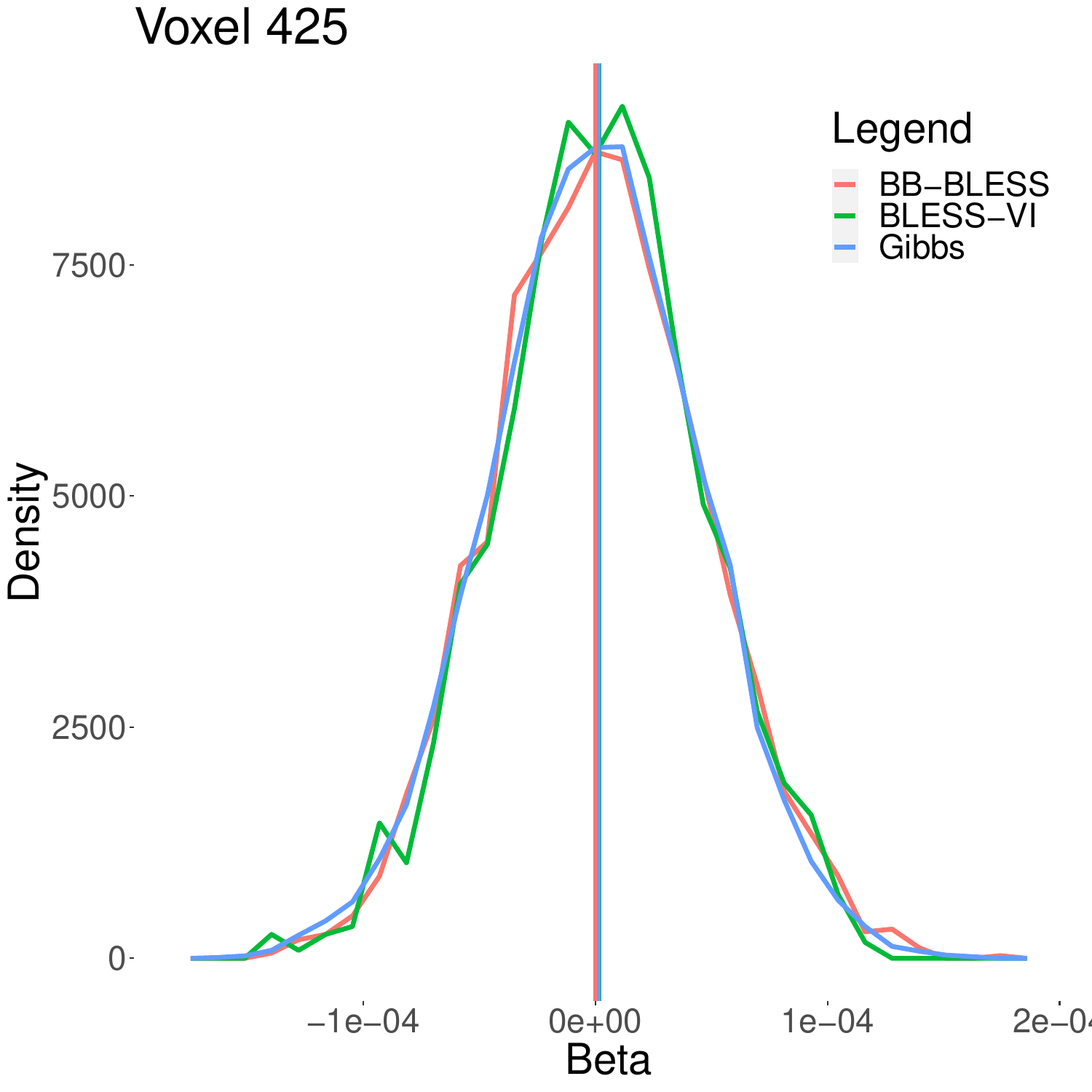} 
\end{subfigure}
\begin{subfigure}{0.24\textwidth}
\caption{Active Voxel}
\includegraphics[width=1\linewidth]{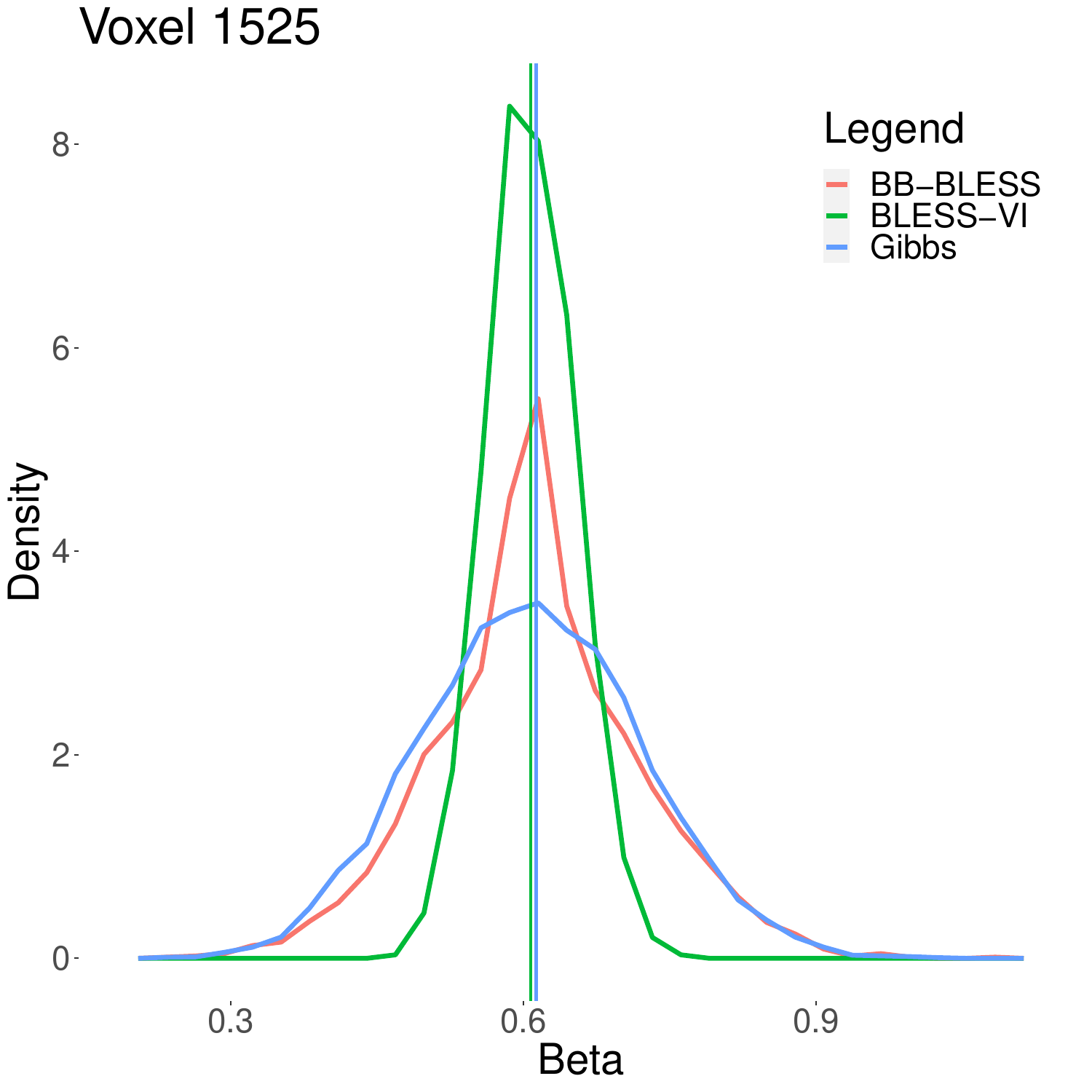} 
\end{subfigure}
\begin{subfigure}{0.24\textwidth}
\caption{KL-divergence}
\includegraphics[width=1\linewidth]{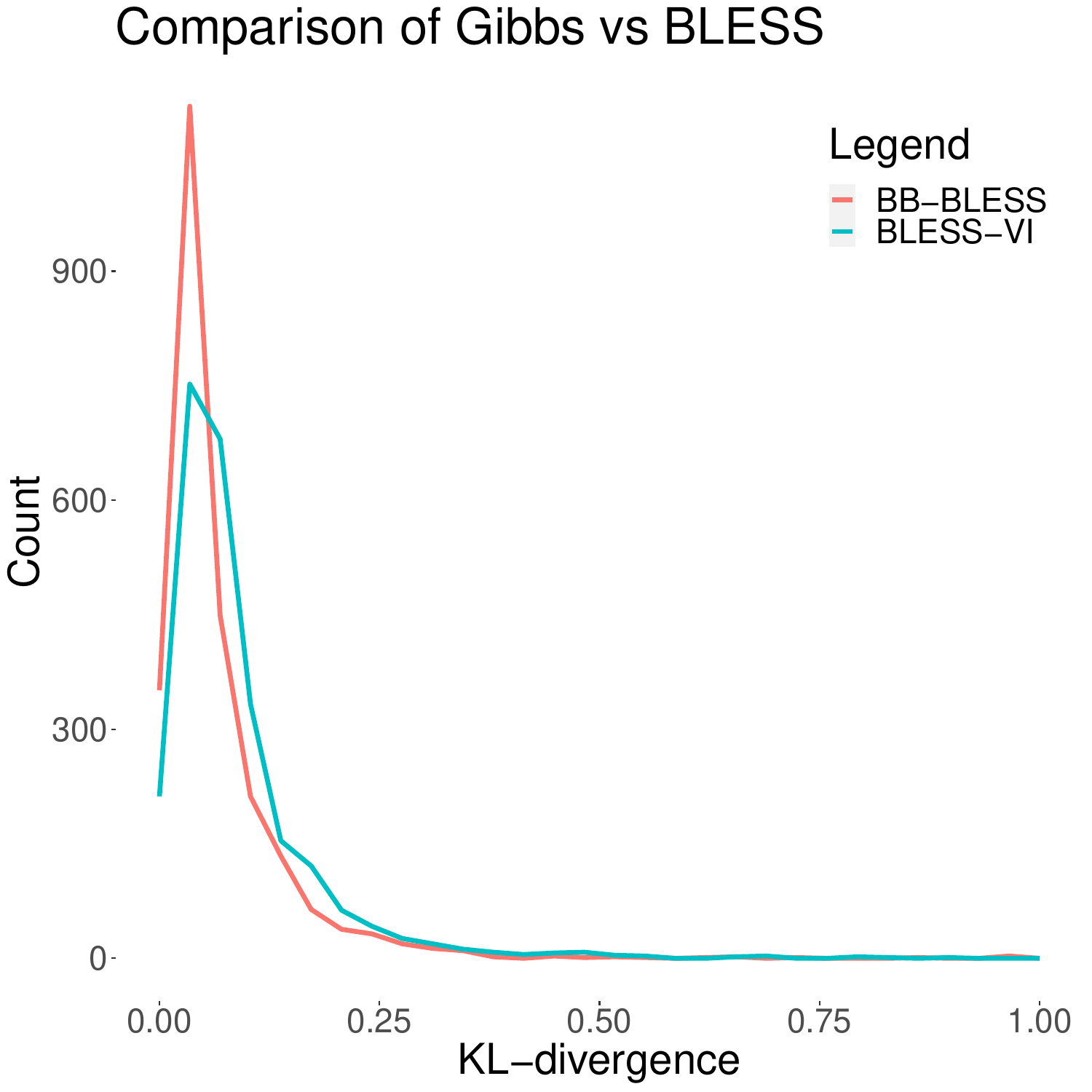} 
\end{subfigure}
\begin{subfigure}{0.24\textwidth}
\caption{Wasserstein metric}
\includegraphics[width=1\linewidth]{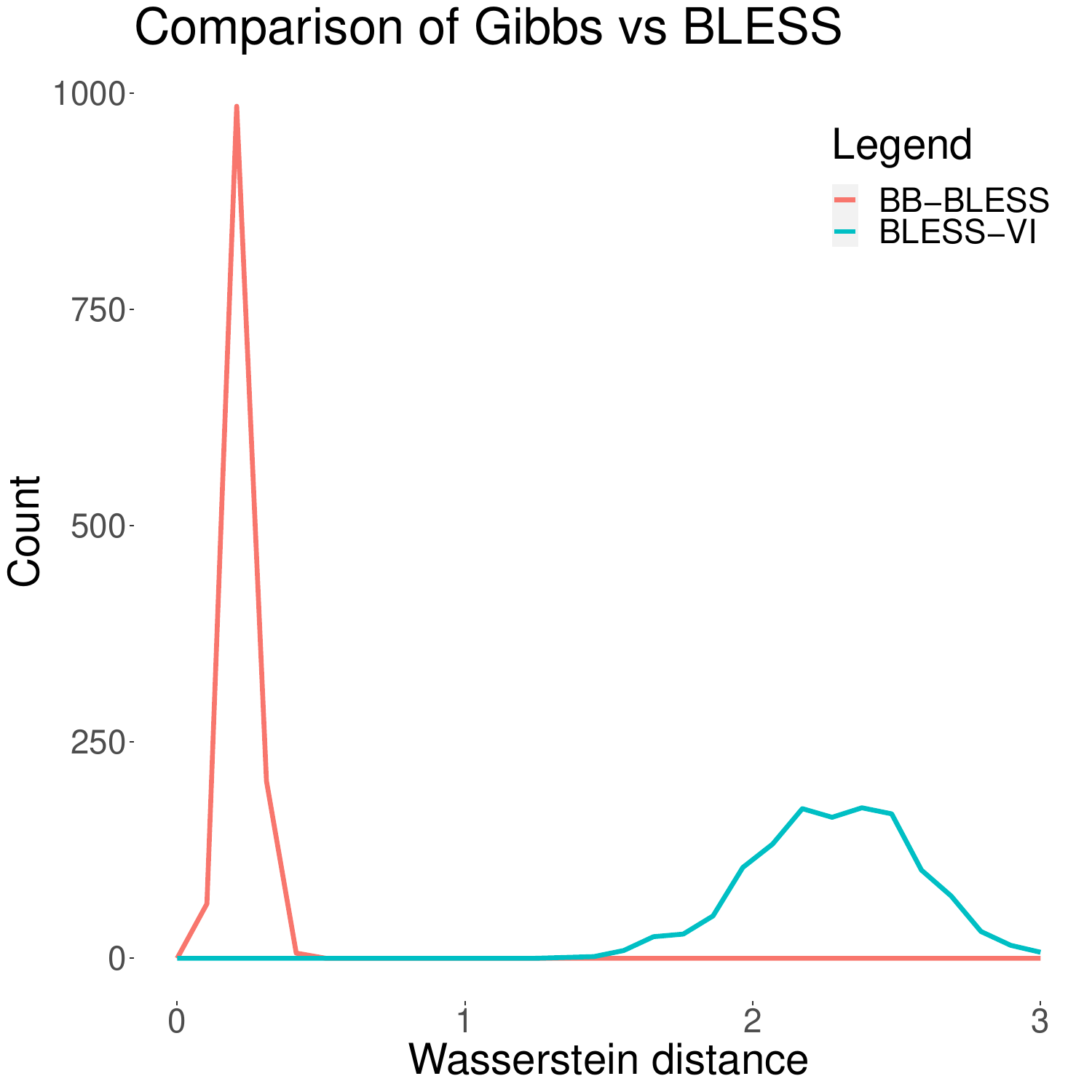}
\end{subfigure}

\begin{subfigure}{0.24\textwidth}
\caption{Posterior Mean: \\Gibbs vs BB-BLESS}
\includegraphics[width=1\linewidth]{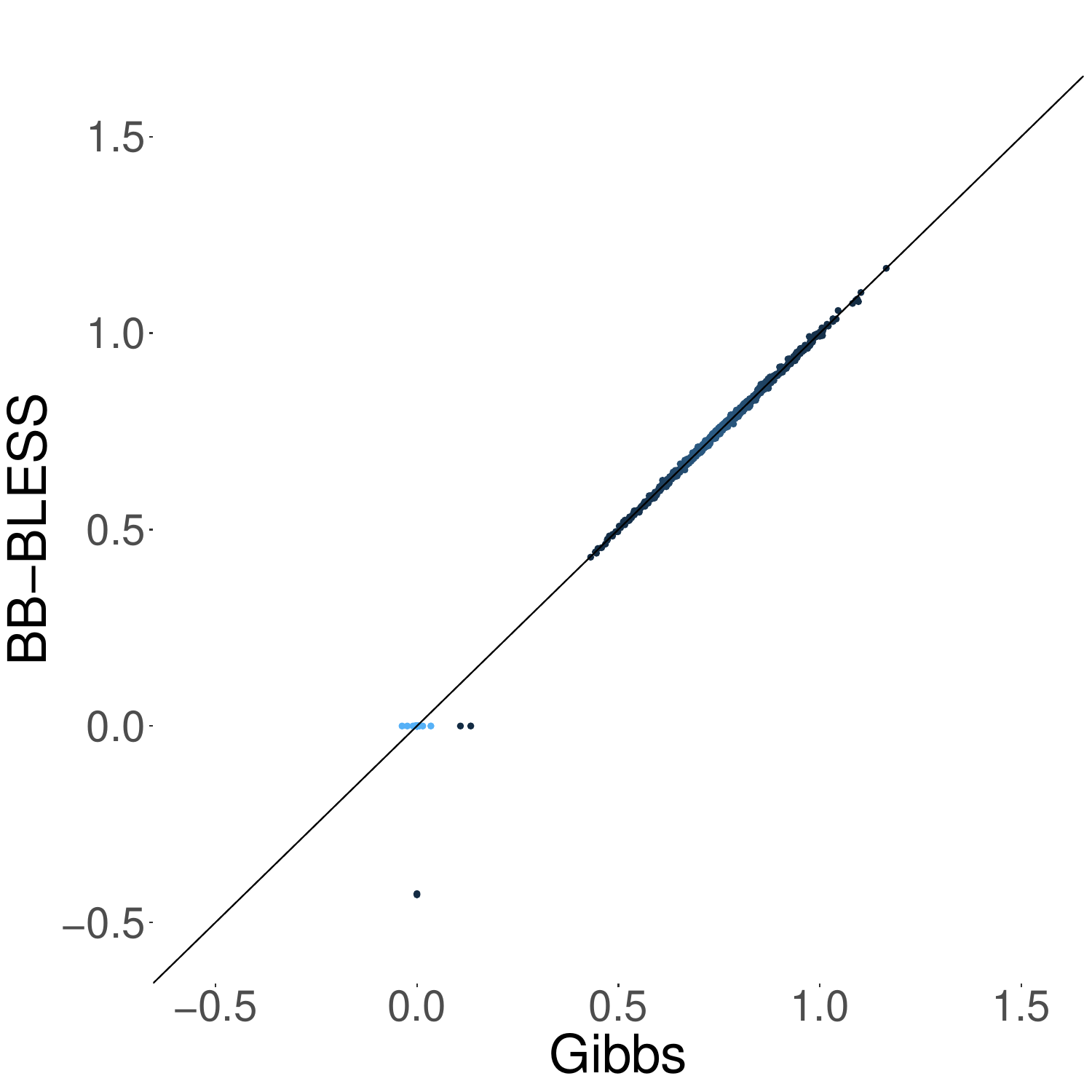} 
\end{subfigure}
\begin{subfigure}{0.24\textwidth}
\caption{Posterior Mean: \\Gibbs vs BLESS-VI}
\includegraphics[width=1\linewidth]{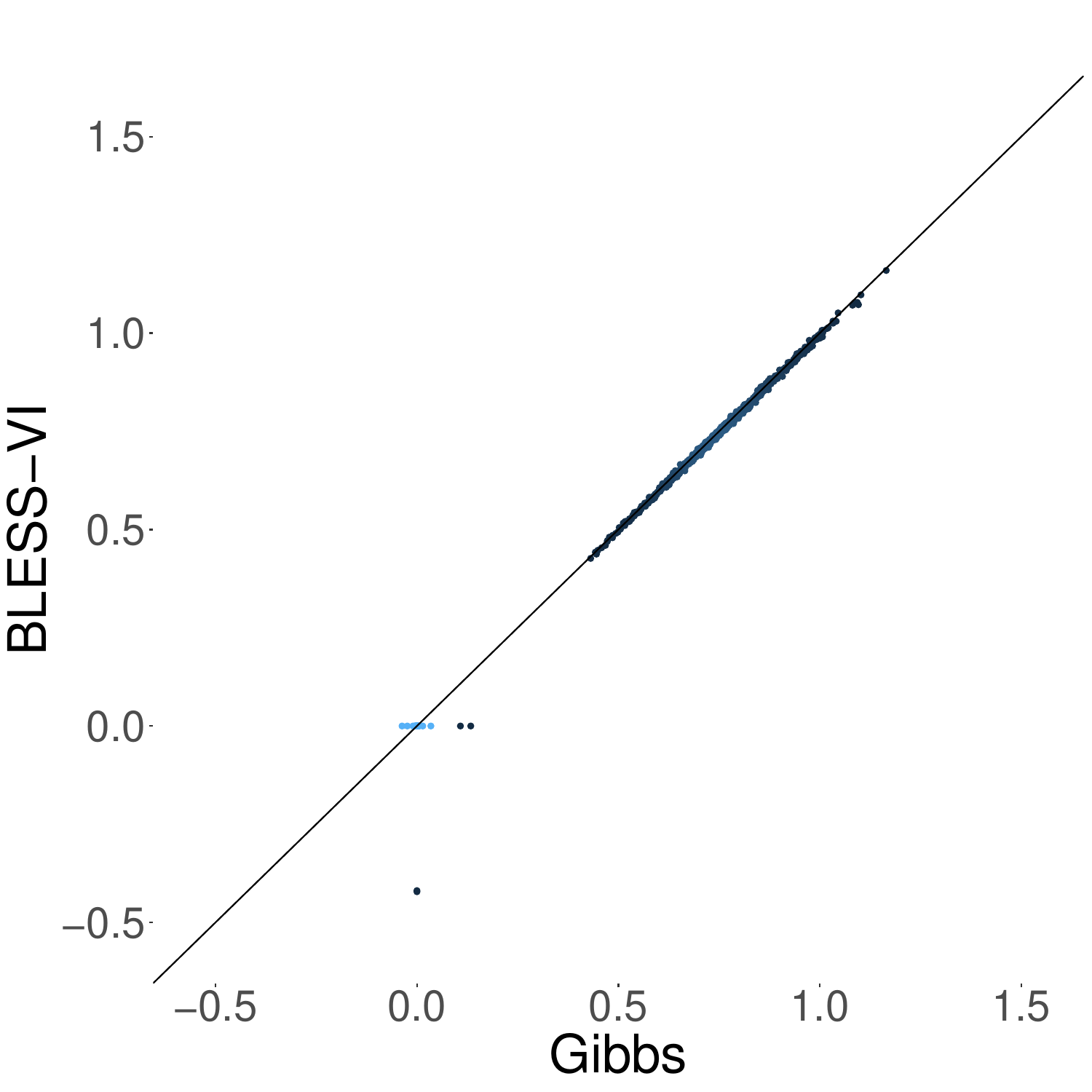}
\end{subfigure}
\begin{subfigure}{0.24\textwidth}
\caption{Posterior Sd: \\Gibbs vs BB-BLESS}
\includegraphics[width=1\linewidth]{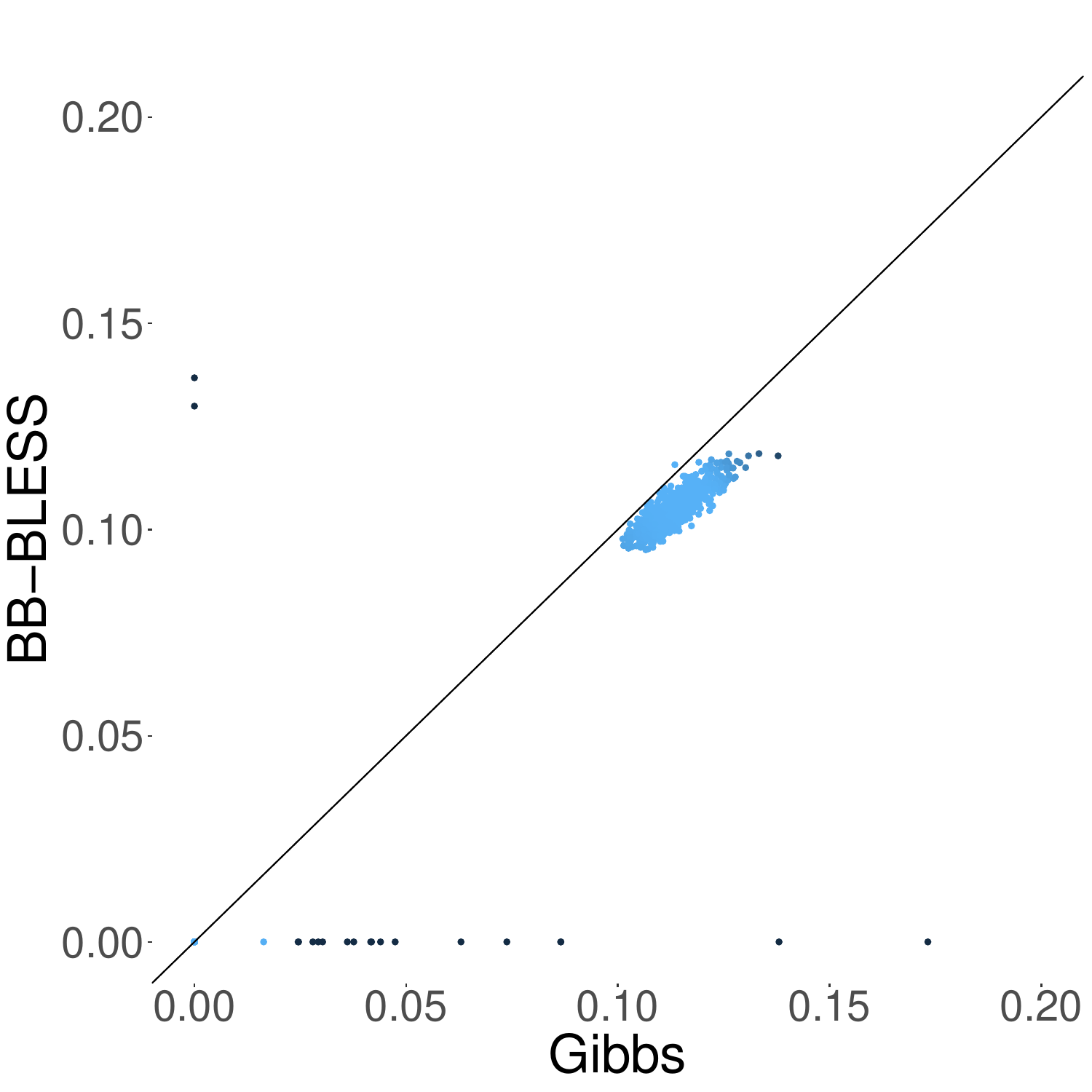}
\end{subfigure}
\begin{subfigure}{0.24\textwidth}
\caption{Posterior Sd: \\Gibbs vs BLESS-VI}
\includegraphics[width=1\linewidth]{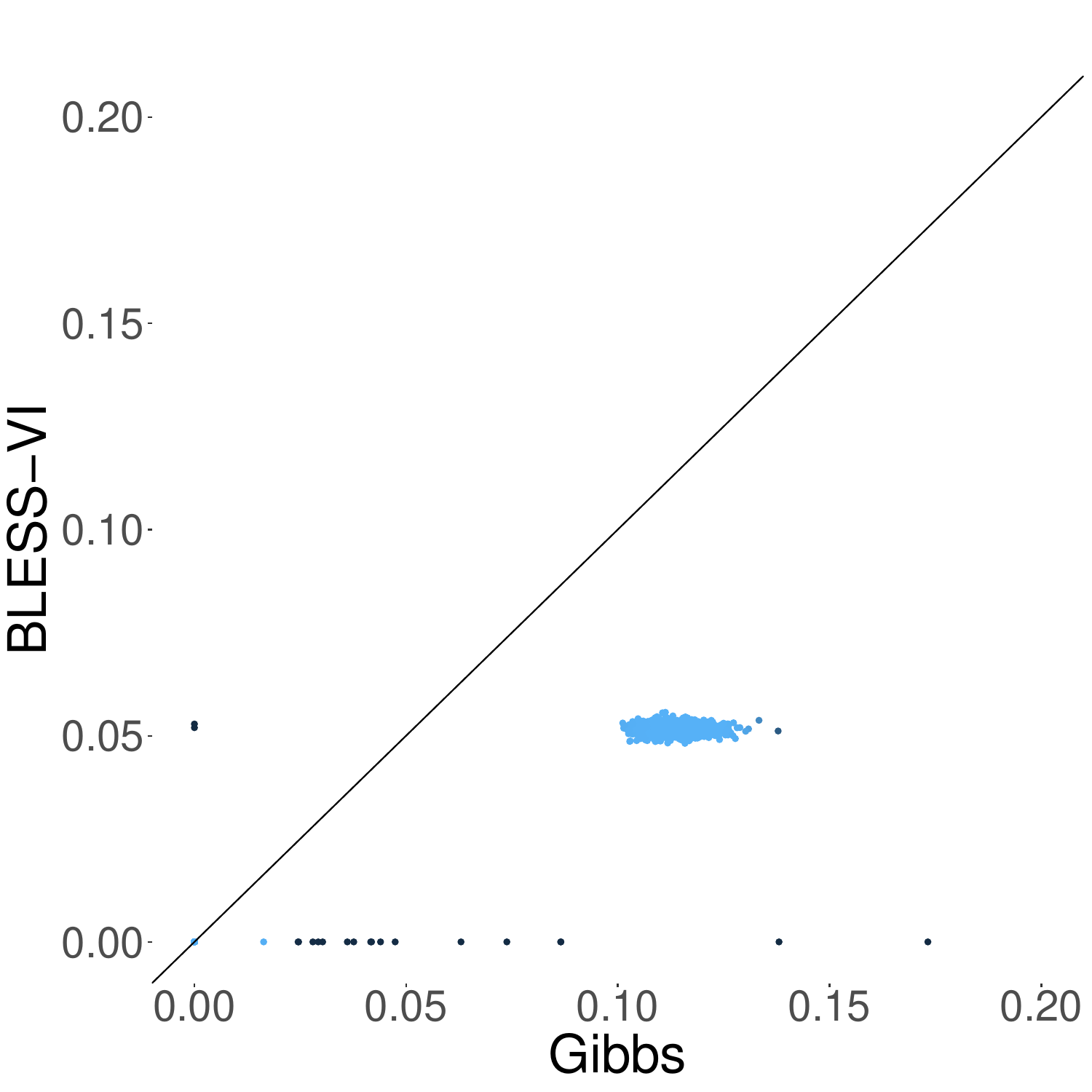}
\end{subfigure}

\caption{Comparison of marginal posterior distributions for an (a) active and (b) inactive voxel between BB-BLESS, BLESS-Gibbs, and BLESS-VI where the posterior mean is indicated via a vertical line. Overall evaluation of marginal posterior distributions for all voxels between Gibbs and BB-BLESS and BLESS-VI via (c) KL-divergence and (d) Wasserstein distance. Comparison of posterior quantities, such as posterior mean (e)-(f) and standard deviation (g)-(h), of the parameter estimates for all voxels for $N=1,000$ and $\lambda=3$ (lighter values indicate higher density of values). Parameters acquired via BLESS-VI exhibit similar point estimates to BB-BLESS and Gibbs but their posterior distributions are too peaked and variances are underestimated.}
\label{fig: bb_bless_mean_etc}
\end{figure}

Firstly, we examine the marginal posterior densities of a random active and inactive voxel. As expected, the posterior variance from BLESS estimated via variational inference is underestimated as the posterior distribution is very peaked around the posterior mean (Fig. \ref{fig: bb_bless_mean_etc}a, b). On the other hand, the posterior estimated via BB-BLESS aligns well with the distribution acquired via the gold standard method of Gibbs sampling. This is further illustrated by comparing the marginal posterior densities of all voxels within an effect image via KL-divergence and Wasserstein distance in Figure \ref{fig: bb_bless_mean_etc}c, d. Both methods show the higher quality of posterior approximation via BB-BLESS compared to BLESS-VI when calculating the discrepancy of the distributions acquired via approximate methods and Gibbs sampling. 

Figure \ref{fig: bb_bless_mean_etc} illustrates that both BB-BLESS and BLESS-VI are able to better capture the posterior mean of all voxel locations within an image when compared to BLESS estimated via Gibbs sampling. However, BLESS-VI severely underestimates the posterior standard deviation for both active and inactive voxels. Lastly, we compare the inference results of our method BLESS-VI, where we use the marginal posterior probability of inclusion as a proxy for inference, to the approximate posterior sampling technique BB-BLESS and the gold standard of BLESS-Gibbs, for which we determine activation via test statistics $t = \hat{\beta} /\sigma_{\hat{\beta}}$, for two simulation study setups. BLESS estimated via Gibbs sampling yields high sensitivity and a very low false positive rate for both settings in Table \ref{tab: inference_BB_bless}. More importantly, the inference results for BB-BLESS and BLESS-VI are very similar, i.e. the false positive rate for both BB-BLESS and BLESS-VI lies at 2.57\% for a sample size of $N=500$ and base rate intensity of $\lambda=1$. Hence, we showcase empirically that, when it comes to inference, thresholding posterior inclusion probabilities in BLESS-VI yields similar results to the approximate posterior sampling approach BB-BLESS which determines effect detection via test statistics. Hence, if a researcher is uninterested in the additional features of BB-BLESS, such as acquiring uncertainty estimates of coefficients or more complex imaging statistics, then the application of BLESS-VI alone can be considered for parameter estimation and inference, as we get empirically similar voxelwise inference results in our simulation studies at a lower overall computational cost for BLESS-VI compared to BB-BLESS. 

\begin{table}[h!]
    \centering
\begin{adjustbox}{width=\columnwidth,center}
    \begin{tabular}{ lcccc|cccc } 
    &\multicolumn{4}{c}{$\bm{N=500}$ \textbf{and} $\bm{\lambda=1}$} &\multicolumn{4}{c}{$\bm{N=1,000}$ \textbf{and} $\bm{\lambda=3}$}\\
     \hline
$\bm{\hat{\beta_1}}$ &\textbf{Bias} & \textbf{Variance} & \textbf{MSE}& &\textbf{Bias} & \textbf{Variance} & \textbf{MSE} \\
  \hline
BLESS-Gibbs & 0.0288 & 0.0406 & 0.0414 & & 0.0060 & 0.0065 & 0.0065 \\ 
  BB-BLESS & -0.0857 & 0.0171 & 0.0245 & &  0.0078 & 0.0063 & 0.0064 \\ 
  BLESS-VI & -0.1037 & \textbf{0.0013} & \textbf{0.0121} & & \textbf{0.0027} & \textbf{0.0010} & \textbf{0.0010}  \\ 
  BSGLMM  & 0.0518 & 0.0120 & 0.0147 &  & 0.0126 & 0.0039 & 0.0040 \\ 
  Firth & \textbf{-0.0182} & 0.0539 & 0.0542 & & \textbf{-0.0027} & 0.0118 & 0.0118 \\ 

   \hline
  $\bm{t_{\hat{\beta_1}}}$ &\textbf{TPR} & \textbf{TDR} & \textbf{FPR}&\textbf{FDR} & \textbf{TPR} & \textbf{TDR} & \textbf{FPR} & \textbf{FDR}\\
  \hline
BLESS-Gibbs & 0.7319 & \textbf{0.9998} & \textbf{0.0001} & \textbf{0.0002} & 0.9999 & \textbf{0.9999} & \textbf{0.0001} & \textbf{0.0001} \\ 
  BB-BLESS & 0.6263 & 0.9606 & 0.0257 & 0.0394 & 0.9999 & 0.9970 & 0.0031 & 0.0030 \\ 
  BLESS-VI & 0.6263 & 0.9606 & 0.0257 & 0.0394 & \textbf{1.0000} & 0.9970 & 0.0031 & 0.0031 \\ 
  BSGLMM & \textbf{0.9991} & 0.9004 & 0.1128 & 0.0996 & \textbf{1.0000} & 0.9027 & 0.1088 & 0.0973 \\ 
  Firth & 0.6566 & 0.8953 & 0.0768 & 0.1047 & \textbf{1.0000} & 0.9637 & 0.0379 & 0.0363  \\ 
   \hline
    \end{tabular}
    \end{adjustbox}
    \caption{Evaluation of parameter estimates and inference results for BLESS (Gibbs, BB, VI), \mbox{BSGLMM} and Firth for two simulation study scenarios ($N=500$, $\lambda = 1$ and $N=1,000$, $\lambda = 3$) for the effect of sex. MSE of parameter estimates for BB-BLESS and BLESS-Gibbs are comparable. Inference results from thresholding posterior inclusion probabilities for BLESS-VI and test statistics for BB-BLESS also achieve similar rates. Run time and complexity analysis can be found in supplements in Section 11.}
    
    \label{tab: inference_BB_bless}
\end{table}

\subsection{BLESS-VI Simulation Study}
\label{sec: sim_study_results_eval}
We extend our simulation study to evaluate the performance of BLESS-VI for a broader set of scenarios with varying sample sizes $N = \{500;\ $1,000$;\ $5,000$\}$, base rate intensities $\lambda = \{1,2,3\}$ and sizes of spatial effect, where $25\%$ (group effect) or $50\%$ (sex effect) of the image are active, compared to BSGLMM and Firth regression. The true and estimated parameter estimates are available in the supplementary materials alongside more results from simulation studies with different spatial priors and varying magnitudes of the slab variance. We will focus on the effect map for the covariate sex and generate 100 datasets for each sample size and base rate scenario to provide robustness by averaging over the results of each dataset. The setup for BLESS-VI is otherwise identical to above.

The quality of parameter estimates and prediction for BLESS-VI, BSGLMM and Firth regression are evaluated via bias, variance and mean squared error (MSE) in Table~\ref{tab: sim_BLESS_bias_params}. BLESS-VI exhibits comparatively low bias for the evaluation of the parameter estimates for the sex effect and moreover outperforms the mass-univariate approach when comparing the quality of the coefficients via MSE. For example, the MSE of the parameter estimates for a small sample size $N=500$ and low base rate intensity $\lambda=1$ is approximately 5 times larger for Firth regression with a value of 0.0563 compared to our method BLESS with a value of 0.0106. This showcases how BLESS adequately regularizes negligible coefficients to zero while the larger effects are unaffected by shrinkage. The quality of the predictive performance is determined by comparing the true empirical lesion rates to the estimated lesion probabilities. BLESS-VI yields slightly better predictive results with respect to MSE, by exhibiting less biased estimates, compared to Firth regression for all scenarios except for the instance with low sample size and base rate intensity ($N=500$, $\lambda=1$) where Firth regression exhibits a slightly lower MSE. This result motivates the usage of BLESS for studies with larger sample sizes where BLESS outperforms the mass-univariate approach. 

Our simulation study enforces 50\% of the voxels as active on the right side of an image for the covariate sex. Hence, by knowing the true location of the effect, we can evaluate the quality of the inference results of BLESS compared to BSGLMM and Firth regression. Effect detection for BLESS is determined by utilizing the latent variables $\bm{\hat{\gamma}}$, marking voxels~$s_j$ significant if $P(\gamma_p(s_j) = 1 | \bm{y}) > 0.5$. For BSGLMM and Firth regression we acquire test statistics $t = \hat{\beta} / \sigma_{\hat{\beta}}$ and threshold them at a significance level of 5\%. We perform a multiple testing adjustment via a Benjamini-Hochberg procedure \citep{benjamini1995}. 

\begin{figure}[!ht]
\centering
\begin{subfigure}{0.24\textwidth}
\caption{TPR}
\includegraphics[width=1\linewidth]{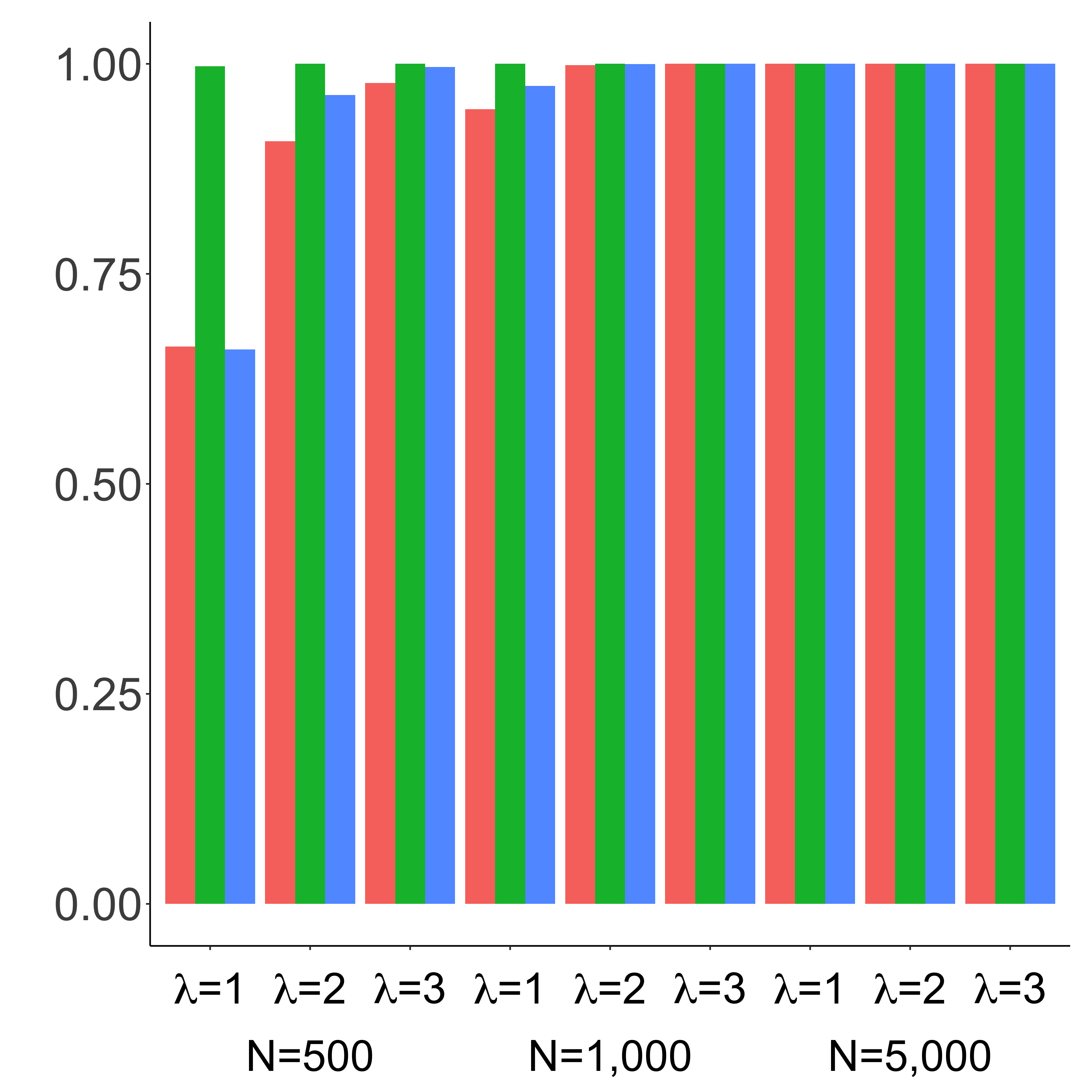}
\end{subfigure}
\begin{subfigure}{0.24\textwidth}
\caption{TDR}
\includegraphics[width=1\linewidth]{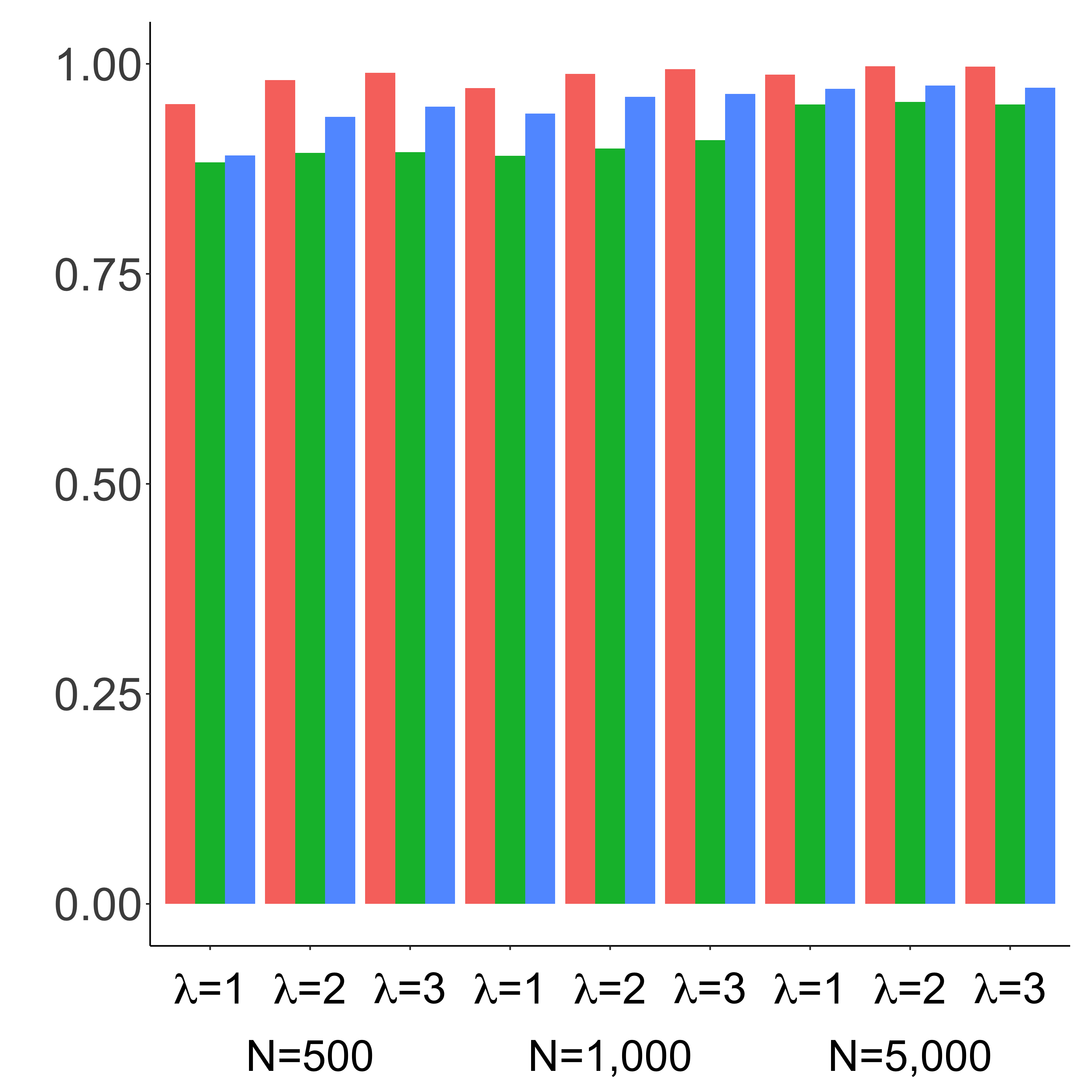}
\end{subfigure}
\begin{subfigure}{0.24\textwidth}
\caption{FPR}
\includegraphics[width=1\linewidth]{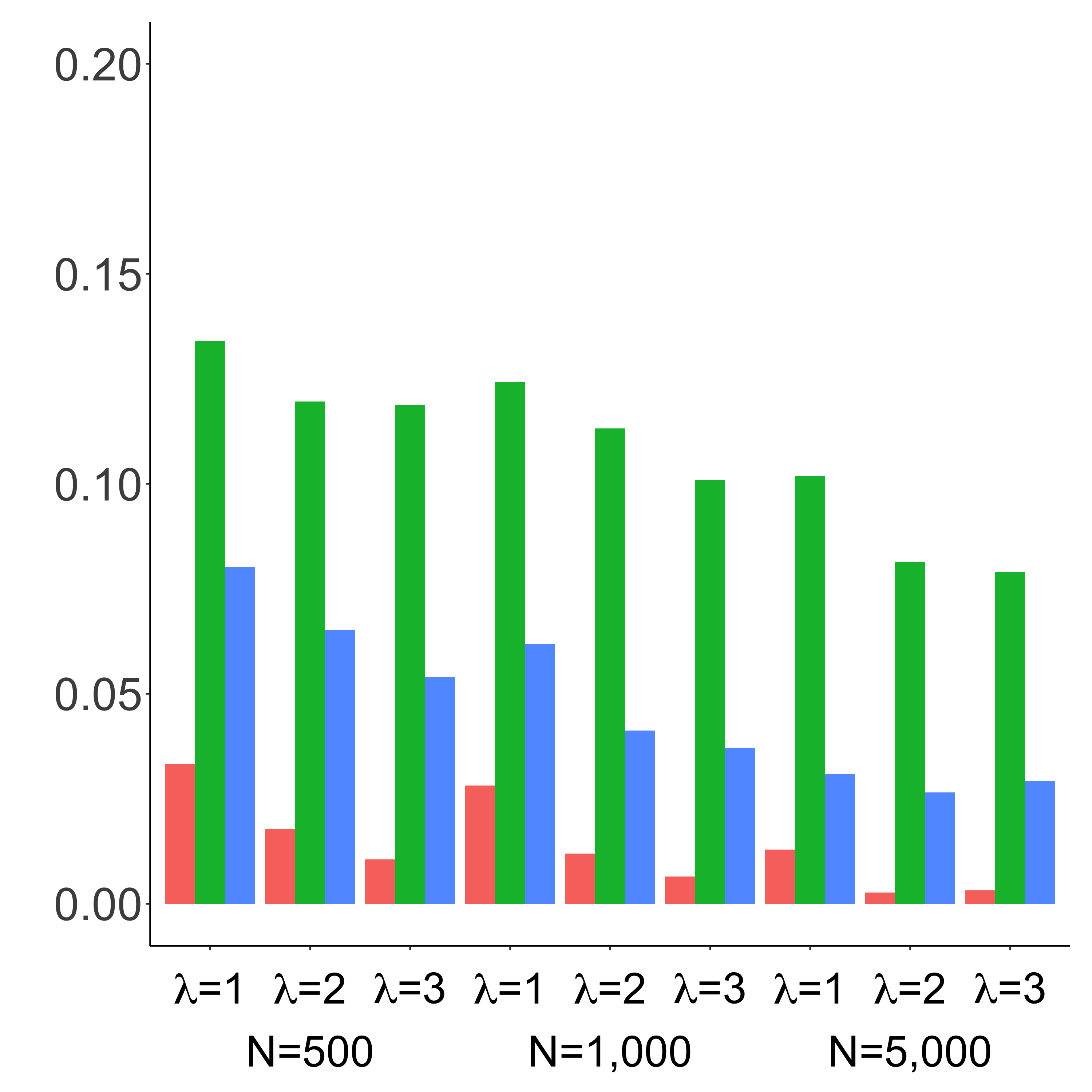}
\end{subfigure}
\begin{subfigure}{0.24\textwidth}
\caption{FDR}
\includegraphics[width=1\linewidth]{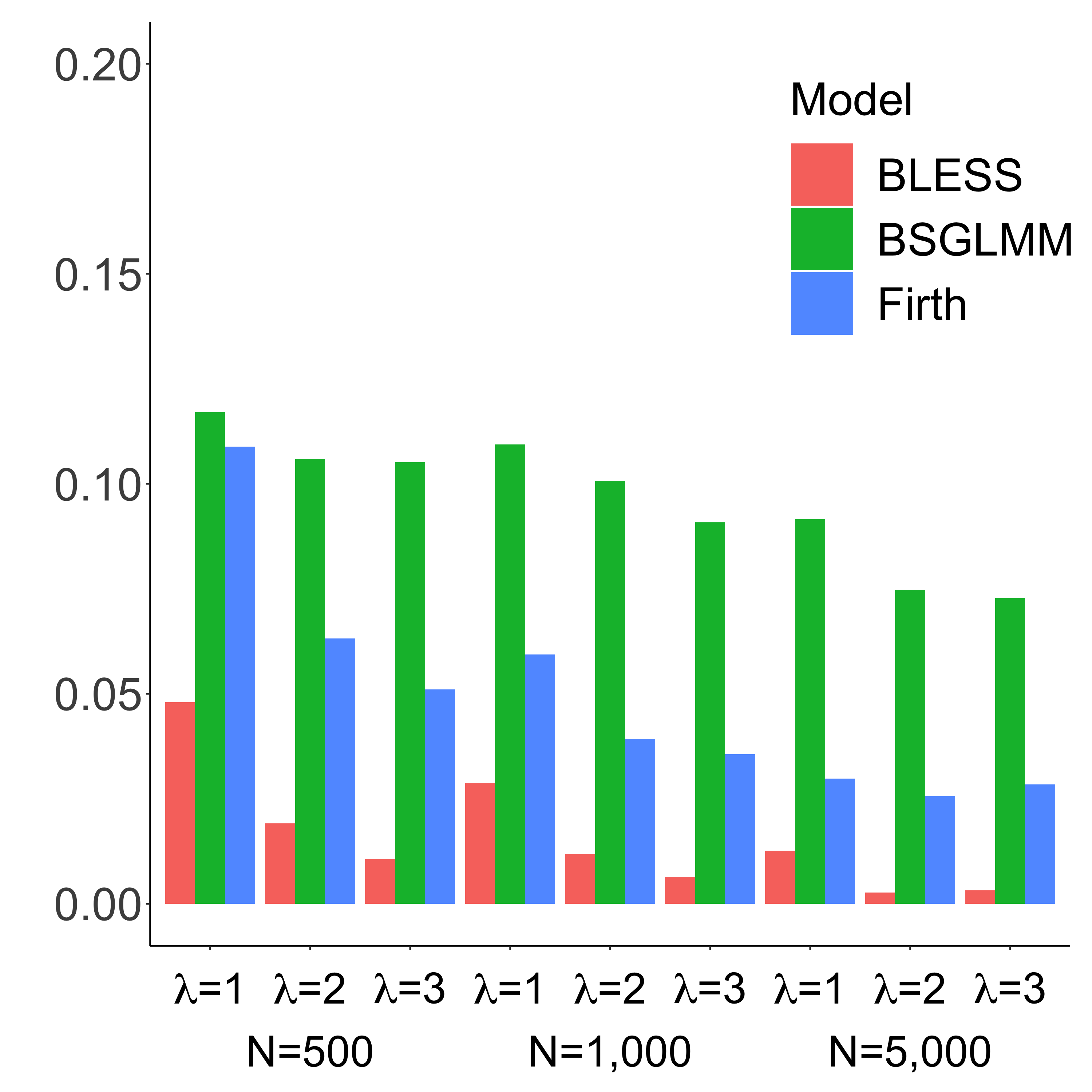}
\end{subfigure}

\caption{Evaluation of inference results from the methods, BLESS, BSGLMM and Firth Regression (FDR correction at 5\%) via True Positive Rate (TPR), True Discovery Rate (TDR), False Positive Rate (FPR) and False Discovery Rate (FDR) for parameter estimate~$\bm{\hat{\beta_1}}$. BLESS outperforms Firth regression and BSGLMM with consistently high TPRs and low FPRs for various sample sizes and base rate intensities.}
\label{fig: inference_sim_study_bless}
\end{figure}

All methods have comparable results with respect to their performance in parameter estimation and prediction. However, the evaluation of the inference results in Figure \ref{fig: inference_sim_study_bless} showcases that the Bayesian spatial model BSGLMM has a particularly high number of false positives and hence a very low level of specificity compared to the other methods. BLESS's key advantage is therefore shown by comparable levels of sensitivity and high values of specificity for all configurations of sample size and base rate intensity. 

\begin{table}[h!]
    \centering
    \begin{adjustbox}{width=1.0\columnwidth,center}
    \begin{tabular}{ lccc|ccc|ccc } 
    \textbf{Parameter Estimate:} $\hat{\bm{\beta}}_1$ &\multicolumn{3}{c}{\textbf{Bias}} &\multicolumn{3}{c}{\textbf{Variance}}&\multicolumn{3}{c}{\textbf{MSE}} \\
  \hline
        \textbf{N=500} & $\lambda=1$ & $\lambda=2$ & $\lambda=3$  & $\lambda=1$ & $\lambda=2$ & $\lambda=3$ & $\lambda=1$ & $\lambda=2$ & $\lambda=3$\\
  \hline
BLESS & -0.0961 & -0.0237 & \textbf{-0.0009} & \textbf{0.0014} & \textbf{0.0019} & \textbf{0.0020} & \textbf{0.0106} & \textbf{0.0024} & \textbf{0.0020} \\ 
  BSGLMM & 0.0280 & 0.0129 & 0.0130 & 0.0117 & 0.0080 & 0.0067 & 0.0125 & 0.0082 & 0.0068 \\ 
  Firth & \textbf{0.0068} & \textbf{-0.0024} & 0.0017 & 0.0562 & 0.0348 & 0.0272 & 0.0563 & 0.0348 & 0.0272 \\ 
    \hline
        \textbf{N=1,000} & $\lambda=1$ & $\lambda=2$ & $\lambda=3$  & $\lambda=1$ & $\lambda=2$ & $\lambda=3$ & $\lambda=1$ & $\lambda=2$ & $\lambda=3$\\
  \hline
   BLESS & -0.0031 & 0.0082 & 0.0019 & \textbf{0.0010} & \textbf{0.0010} & \textbf{0.0010} & \textbf{0.0010} & \textbf{0.0011} & \textbf{0.0010} \\ 
  BSGLMM & 0.0127 & 0.0106 & 0.0066 & 0.0063 & 0.0045 & 0.0039 & 0.0064 & 0.0046 & 0.0039 \\ 
  Firth & \textbf{-0.0002} & \textbf{0.0026} & \textbf{0.0005} & 0.0271 & 0.0171 & 0.0135 & 0.0271 & 0.0171 & 0.0135 \\ 

    \hline
        \textbf{N=5,000} & $\lambda=1$ & $\lambda=2$ & $\lambda=3$  & $\lambda=1$ & $\lambda=2$ & $\lambda=3$ & $\lambda=1$ & $\lambda=2$ & $\lambda=3$\\
  \hline
  BLESS & 0.0032 & 0.0039 & -0.0011 & \textbf{0.0002} & \textbf{0.0002} & \textbf{0.0002} & \textbf{0.0002} & \textbf{0.0002} & \textbf{0.0002} \\ 
  BSGLMM & 0.0057 & 0.0054 & \textbf{-0.0006} & 0.0018 & 0.0014 & 0.0012 & 0.0018 & 0.0014 & 0.0012 \\ 
  Firth & \textbf{0.0022} & \textbf{0.0031} & -0.0023 & 0.0053 & 0.0034 & 0.0027 & 0.0053 & 0.0034 & 0.0027 \\ 
\hline
   &\\
   \textbf{Predictive Performance:} $\bm{\hat{y}}$ &\multicolumn{3}{c}{\textbf{Bias}} &\multicolumn{3}{c}{\textbf{Variance}}&\multicolumn{3}{c}{\textbf{MSE}} \\
  \hline
        \textbf{N=500} & $\lambda=1$ & $\lambda=2$ & $\lambda=3$  & $\lambda=1$ & $\lambda=2$ & $\lambda=3$ & $\lambda=1$ & $\lambda=2$ & $\lambda=3$\\
  \hline
BLESS & -0.0078 & -0.0052 & -0.0032 & 0.0011 & 0.0017 & 0.0018 & 0.0022 & 0.0031 & 0.0034 \\ 
  BSGLMM & \textbf{-0.0027} & \textbf{-0.0025} & \textbf{-0.0020} & \textbf{0.0002} & \textbf{0.0004} & \textbf{0.0007} & \textbf{0.0002} & \textbf{0.0004} & \textbf{0.0007} \\ 
  Firth & 0.0170 & 0.0140 & 0.0122 & 0.0009 & 0.0016 & 0.0022 & 0.0018 & 0.0032 & 0.0043 \\ 
  \hline
        \textbf{N=1,000} & $\lambda=1$ & $\lambda=2$ & $\lambda=3$  & $\lambda=1$ & $\lambda=2$ & $\lambda=3$ & $\lambda=1$ & $\lambda=2$ & $\lambda=3$\\
  \hline
  BLESS & -0.0015 & \textbf{0.0007} & \textbf{-0.0013} & 0.0004 & 0.0005 & 0.0007 & 0.0008 & 0.0010 & 0.0012 \\ 
  BSGLMM & \textbf{-0.0010} & -0.0018 & -0.0020 & \textbf{0.0001} & \textbf{0.0002} & \textbf{0.0003} & \textbf{0.0001} & \textbf{0.0002} & \textbf{0.0003} \\ 
  Firth & 0.0082 & 0.0082 & 0.0056 & 0.0005 & 0.0008 & 0.0011 & 0.0009 & 0.0016 & 0.0021 \\ 
  \hline
        \textbf{N=5,000} & $\lambda=1$ & $\lambda=2$ & $\lambda=3$  & $\lambda=1$ & $\lambda=2$ & $\lambda=3$ & $\lambda=1$ & $\lambda=2$ & $\lambda=3$\\
  \hline
 BLESS & -0.0006 & \textbf{0.0000} & 0.0008 & 0.0001 & \textbf{0.0001} & \textbf{0.0001} & 0.0001 & 0.0002 & 0.0003 \\ 
  BSGLMM & \textbf{-0.0004} & -0.0008 & \textbf{-0.0001} & \textbf{0.0000} & \textbf{0.0001} & \textbf{0.0001} & \textbf{0.0000} & \textbf{0.0001} & \textbf{0.0001} \\ 
  Firth & 0.0010 & 0.0015 & 0.0020 & 0.0001 & 0.0002 & 0.0002 & 0.0002 & 0.0003 & 0.0004 \\ 
   \hline
    \end{tabular}
    \end{adjustbox}
    \caption{Evaluation of parameter estimates from the methods, BLESS-VI, BSGLMM and Firth Regression via bias, variance and MSE of the spatially-varying coefficients $\bm{\hat{\beta_1}}$, and the predictive performance $\bm{\hat{y}}$. Improved bias and MSE for prediction for BLESS compared to Firth regression due to selective shrinkage property of BLESS.}
    \label{tab: sim_BLESS_bias_params}
\end{table}

\section{UK Biobank Application}
\label{sec: UKBB}
\subsection{Data Description and Model Estimation}
\label{sec: ukbb_description}
Our motivating data set is from the UK Biobank, a large-scale biomedical database containing imaging data from predominately healthy individuals. With a target of 100,000 subjects, there is currently imaging data available for 40,000 participants \citep{miller2016}. We refer the reader to \cite{miller2016} for a detailed description of the scanning and processing protocols. Our goal is to map the influence of risk factors on the incidence of matter hyperintensities to understand their potential clinical significance and how they may contribute to neurological and cognitive deficits.  Our data set consists of $N =$\ 38,331 subjects for which white matter hyperintensity binary lesion masks have been generated via the automatic lesion segmentation algorithm BIANCA \citep{griffanti2016}. The binary lesion maps in subject space are then registered to a common 2mm MNI template across subjects. Each 3D binary image with voxel size $2\times2\times2$ mm$^3$ and dimensions $91\times109\times91$ contains a total of 902,629 voxel locations. Our region of interest lies in the white matter tracts of the brain and hence the total number of voxels is restricted to $M =$\ 54,728 by masking the 3D-lesion masks. We are interested in modeling the influence of age on lesion incidence while accounting for the confounding variables sex, head size scaling factor and the interaction of age and sex. In order to ensure interpretability across studies we have chosen confounds based on research by \cite{alfaro-almagro2021} where a head size scaling factor is commonly included to normalize brain tissue volumes for head size compared to the MNI template. The mean age of the participants in our study is $63.6$ years ($\pm\ 7.5$ years) and $53.04\%$ of individuals are female (20,332 women). 

For model estimation, we firstly perform backwards dynamic posterior exploration over $\nu_0 = \{\exp(-10), \dots, \exp(-3)\}$ to help with the optimization of the variational parameters; otherwise, we fit the model identically to the simulation study as described in the previous section. The regularization and marginal plot for this application to the UK Biobank can be found in the supplementary material in Section 4.3. We further estimate BB-BLESS by acquiring $B=1,500$ bootstrap replicates in which we re-weight the likelihood by drawing Dirichlet weights for every subject with a concentration parameter $\alpha=1$ and perturb the prior mean of the structured spike-and-slab prior by drawing a ``jitter'' from $\mathcal{N}(0, \nu_0)$. We initialize the parameters via the results from the DPE procedure and validate the behavior of the annealing-like strategy by examining the regularization and marginal plot. This approximate posterior sampling method remains highly scalable as each optimization can be performed in parallel.   
\subsection{Results}
\label{sec: ukbb_results}
\begin{figure}[h!]
\rotatebox[origin=l]{90}{\ \small{(1) Raw Effect Size Image}\ \ \ \ \ \ \ }
\begin{subfigure}{0.3\textwidth}
\caption{BB-BLESS}
\includegraphics[width=1\linewidth]{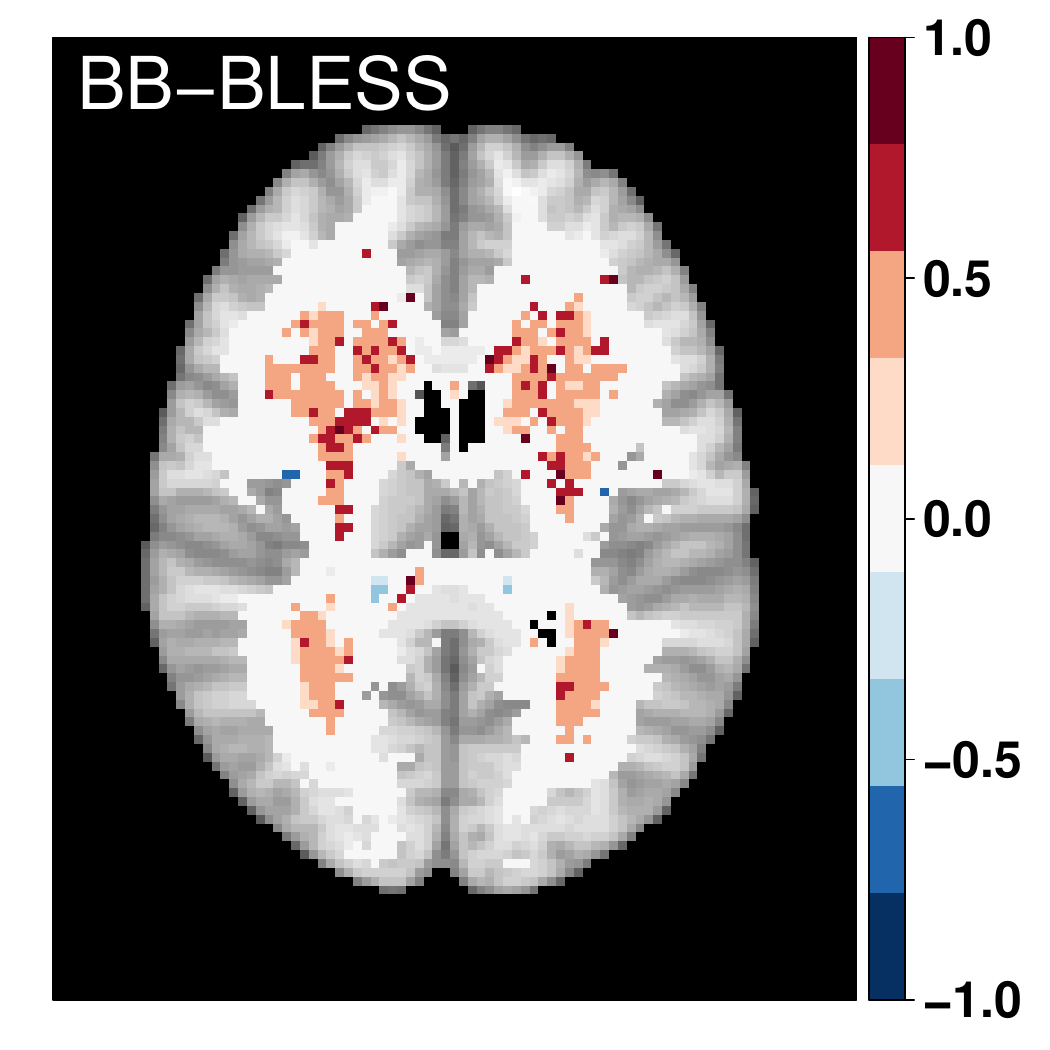}
\end{subfigure}
\begin{subfigure}{0.3\textwidth}
\caption{BLESS-VI}
\includegraphics[width=1\linewidth]{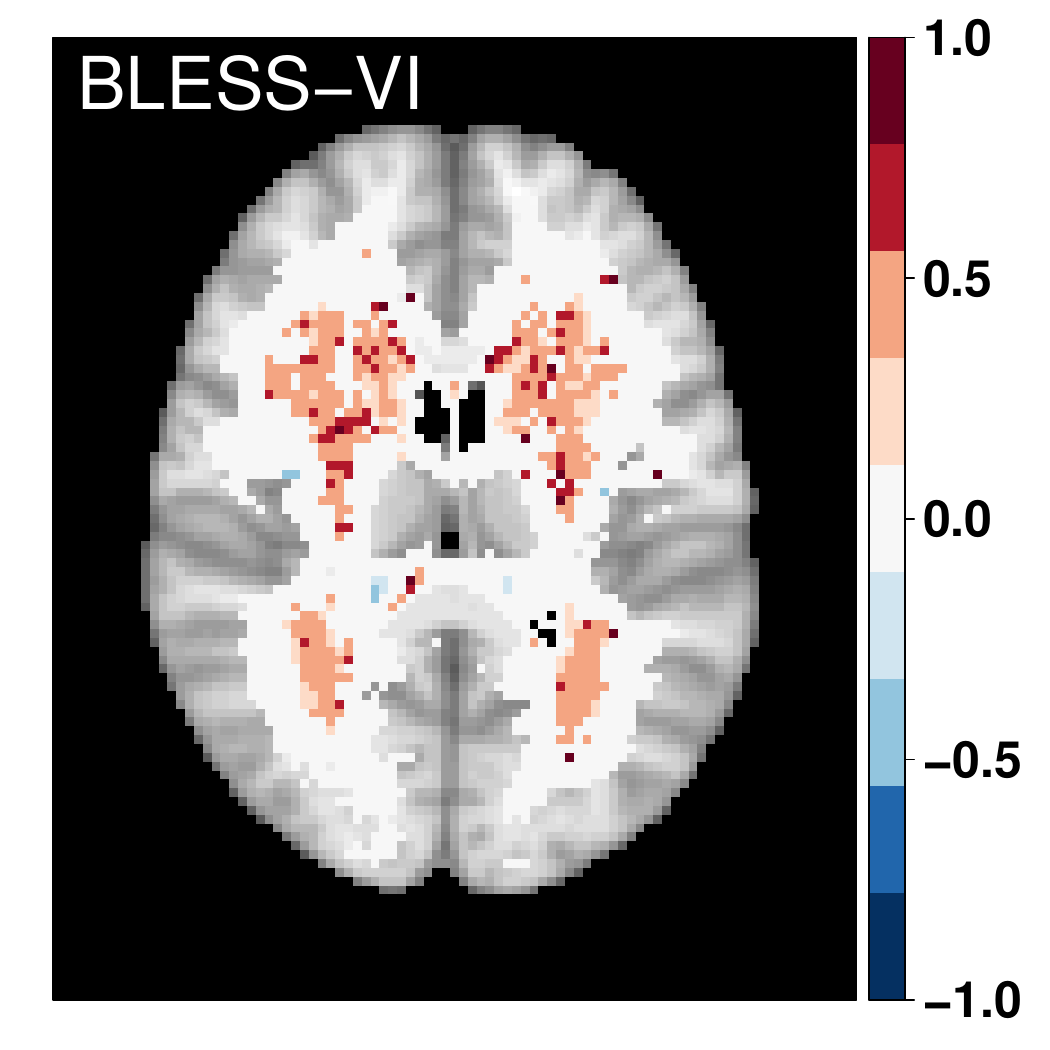}
\end{subfigure}
\begin{subfigure}{0.3\textwidth}
\caption{Firth Regression}
\includegraphics[width=1\linewidth]{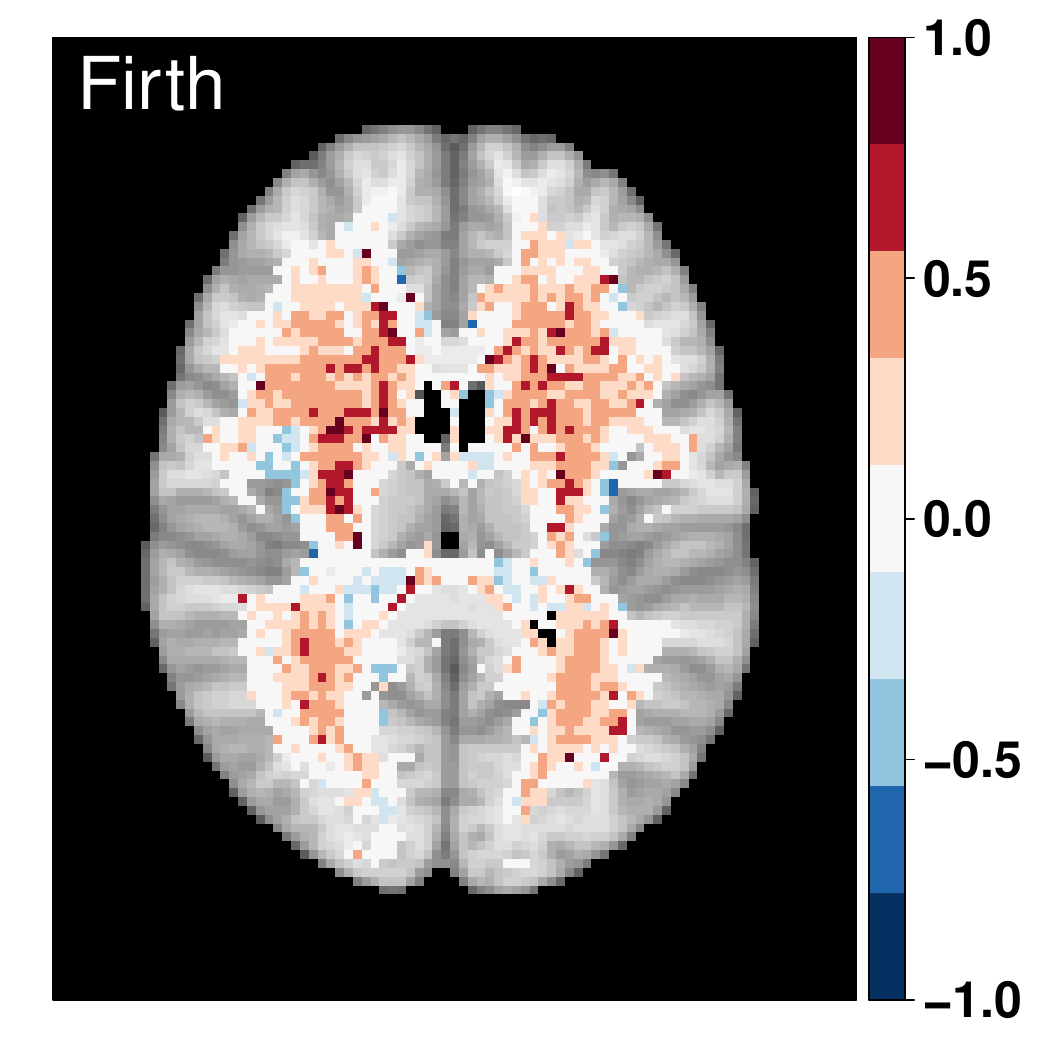}
\end{subfigure}

\rotatebox[origin=l]{90}{\ \ \ \small{(2) Significance Image}}
\begin{subfigure}{0.3\textwidth}
\includegraphics[width=1\linewidth]{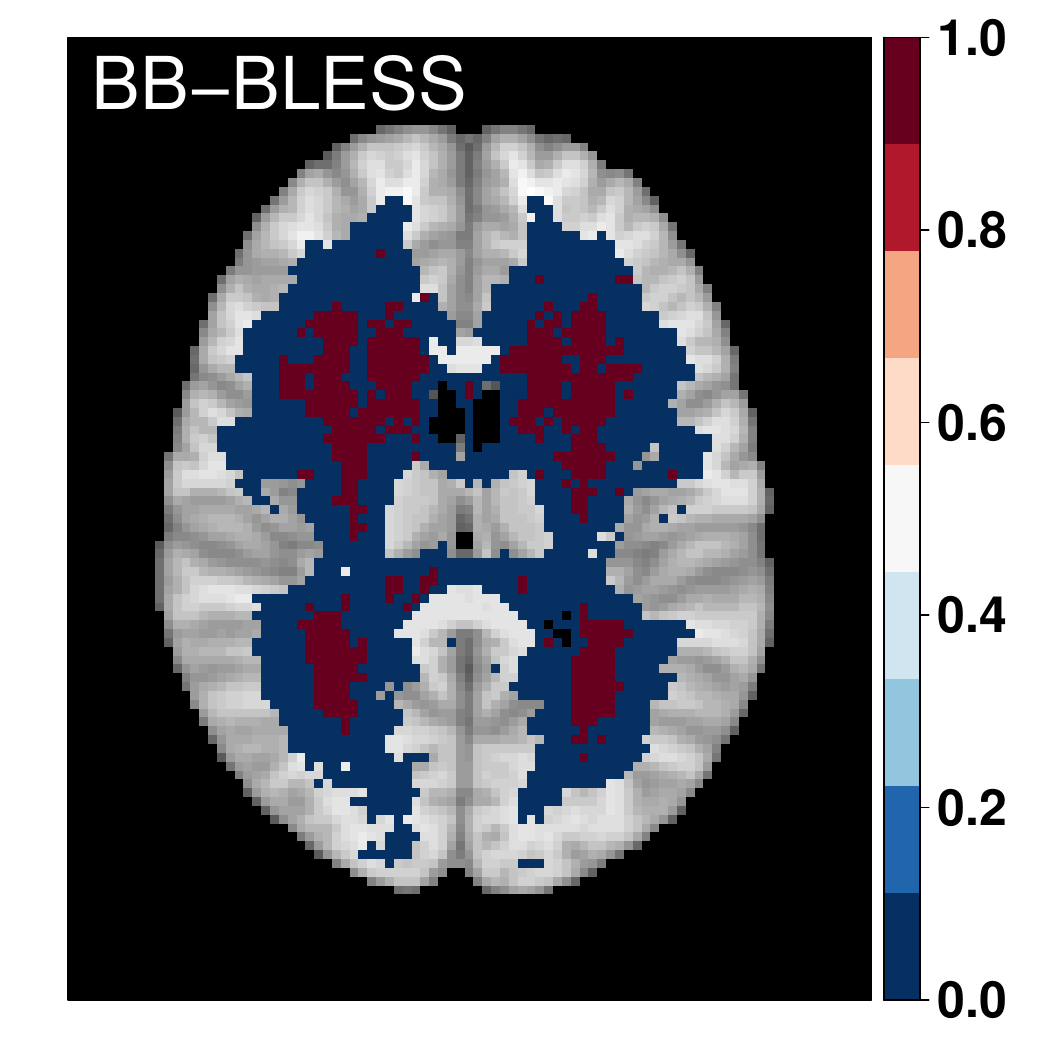}
\end{subfigure}
\begin{subfigure}{0.3\textwidth}
\includegraphics[width=1\linewidth]{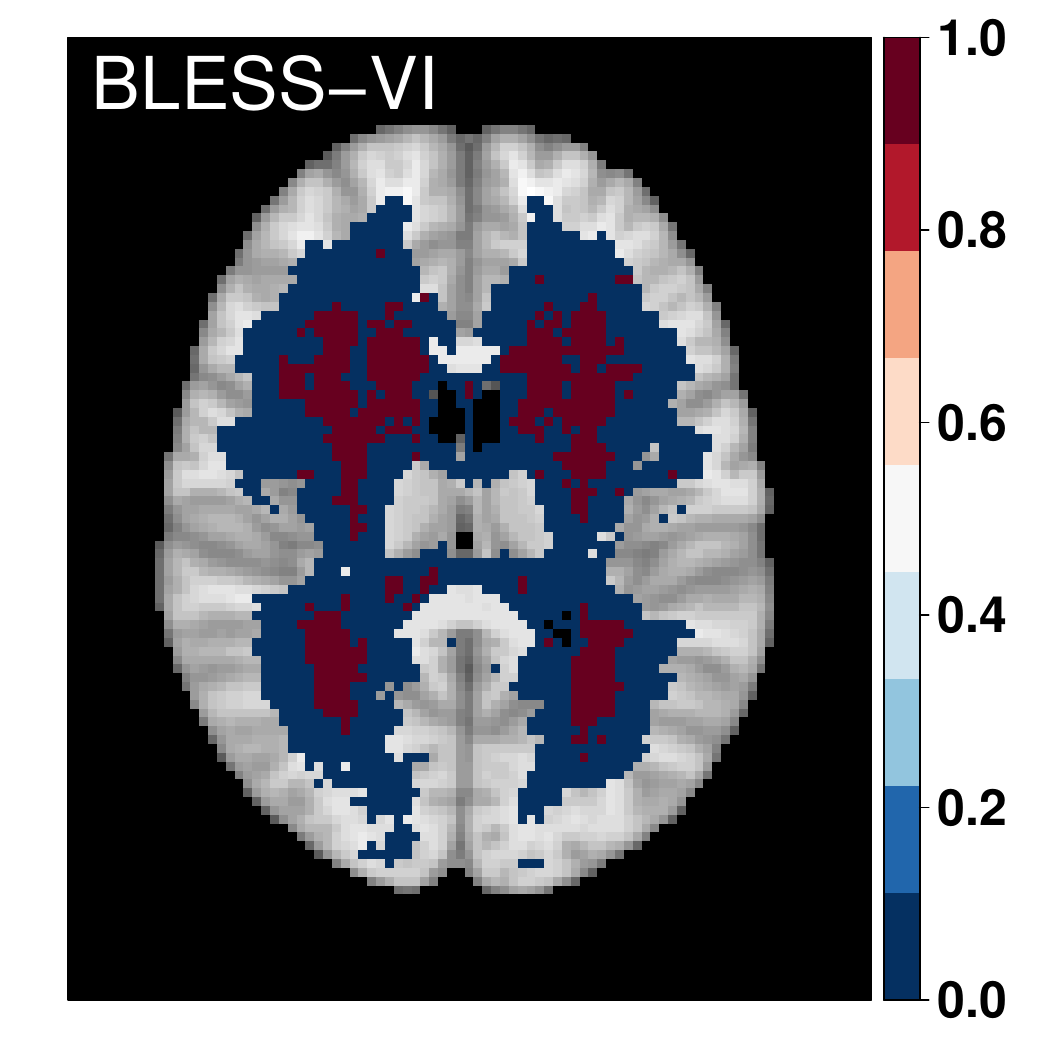}
\end{subfigure}
\begin{subfigure}{0.3\textwidth}
\includegraphics[width=1\linewidth]{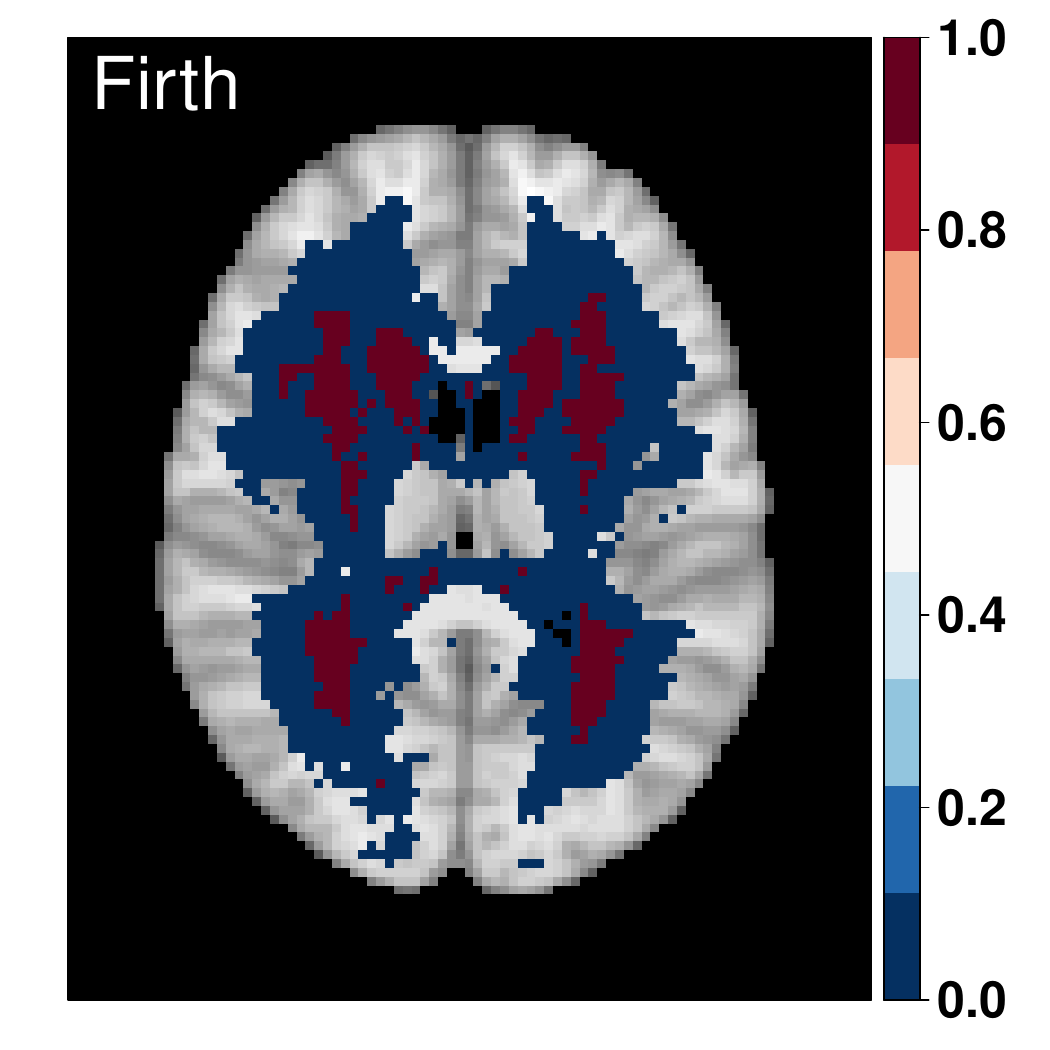}
\end{subfigure}

\rotatebox[origin=l]{90}{\ \ \ \ \ \ \small{(3) Scatterplots}}
\begin{subfigure}{0.3\textwidth}
\includegraphics[width=1\linewidth]{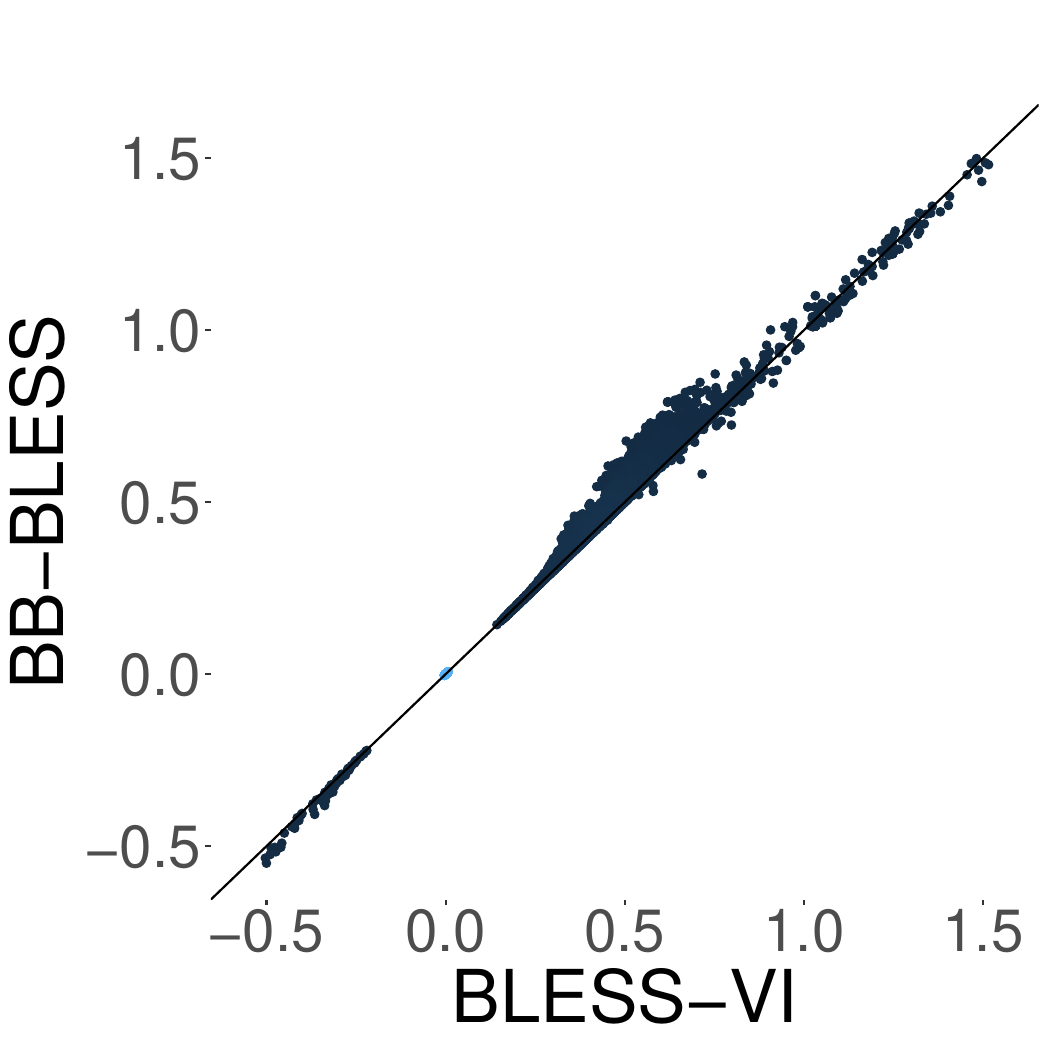}
\end{subfigure}
\begin{subfigure}{0.3\textwidth}
\includegraphics[width=1\linewidth]{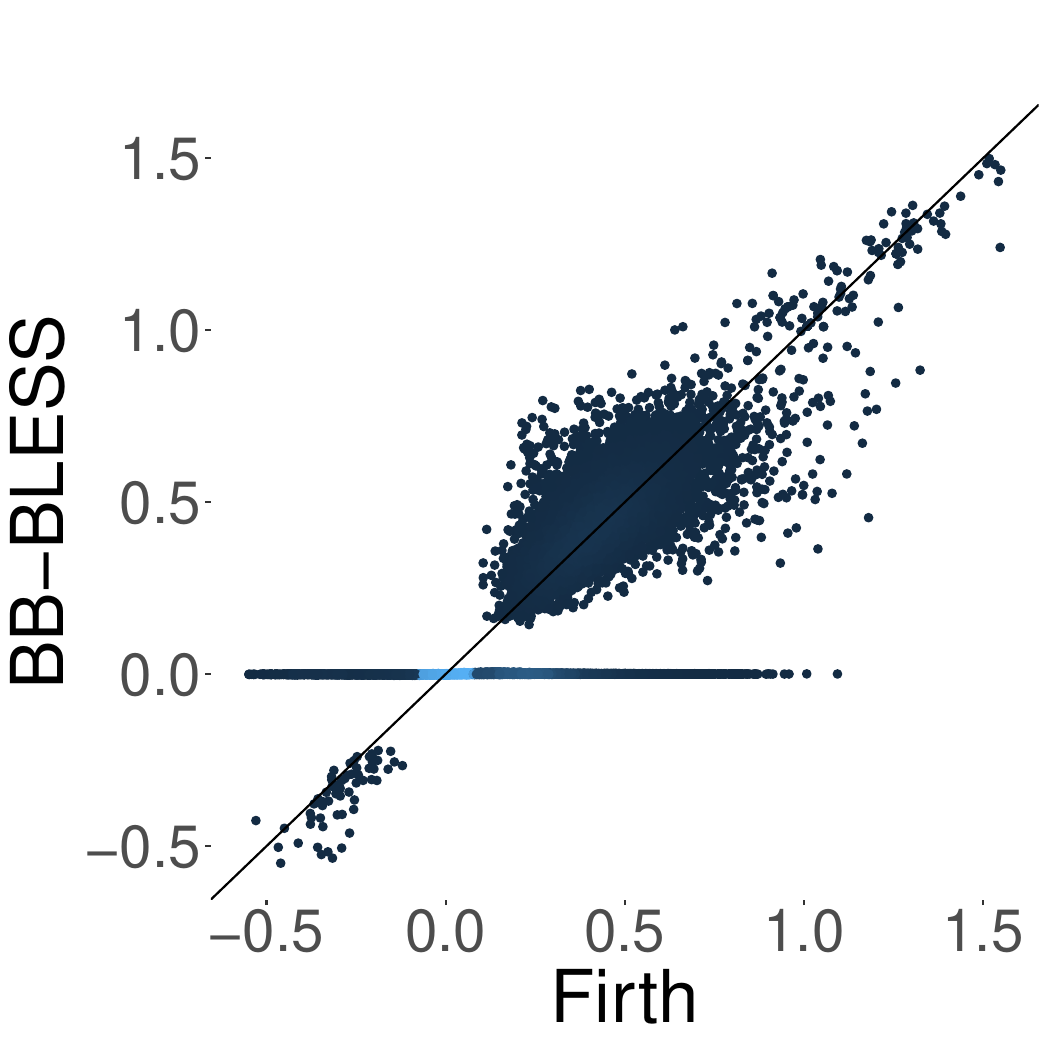}
\end{subfigure}
\begin{subfigure}{0.3\textwidth}
\includegraphics[width=1\linewidth]{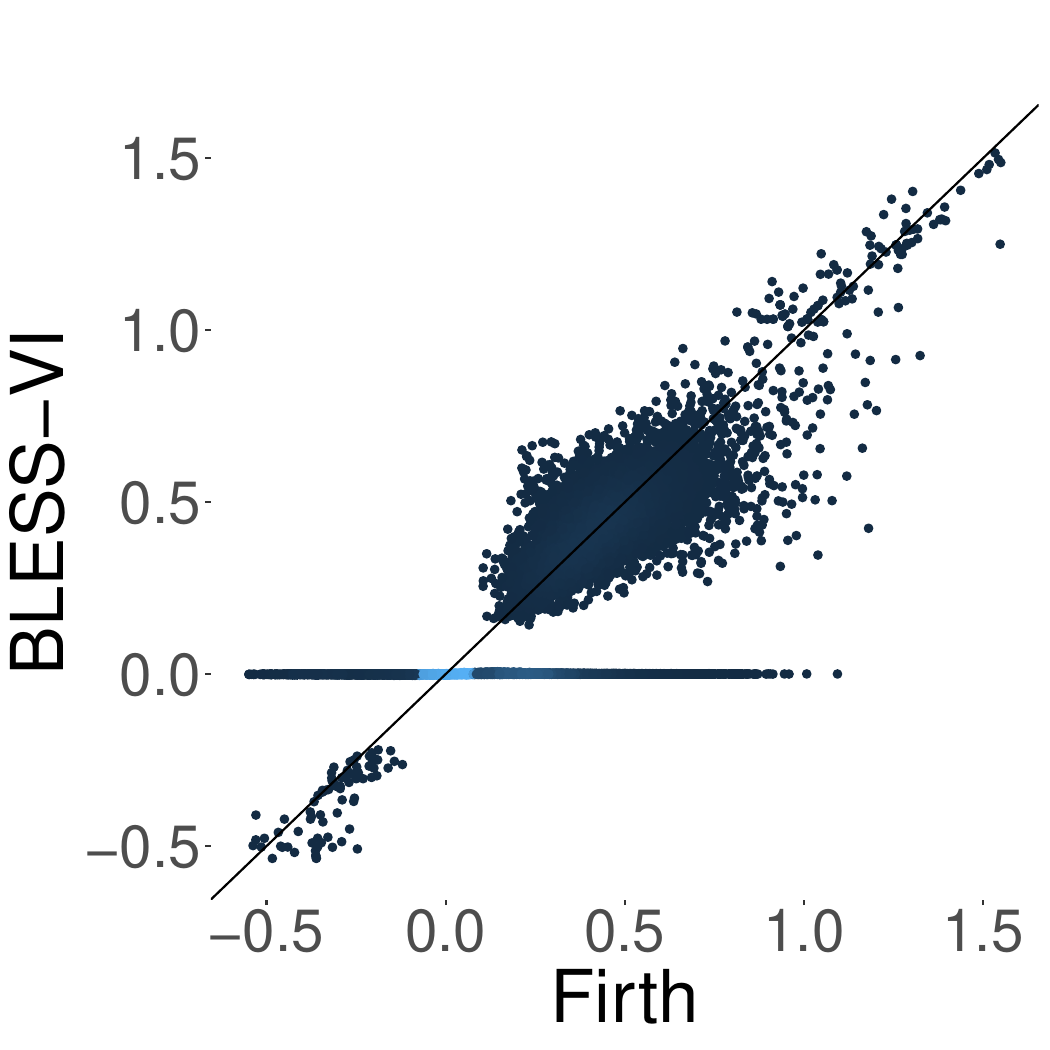}
\end{subfigure}

\caption{Comparison of results between (a) BB-BLESS, (b) BLESS-VI, and (c) Firth Regression for a single axial slice ($z=45$, third dimension of 3D image). (1) Spatially-varying age coefficient maps. (2) Thresholded age significance maps where the threshold for BLESS-VI is determined via the probability of inclusion/exclusion $P(\gamma_p(s_j) | \hat{\bm{\beta}}, \hat{\bm{\theta}}) \geq 0.5$ and the threshold for BB-BLESS and Firth regression via the test statistic $ t = | \hat{\beta}/\hat{\sigma}_{\hat{\beta}} | \geq 1.96 $ (significant voxels: red, not signficant voxels: blue, FDR-correction applied at 5\%). (3) The scatterplots compare the age coefficients for all voxel locations within the 3D image (lighter values indicate higher density of values). The parameter maps estimated via BLESS in (1a) and (1b) exhibit a larger spatial area with values close to 0 compared to Firth in (1c). The scatterplots in (3) show that BLESS regularizes small effects almost completely to 0 compared to Firth.}
\label{fig: ukbb_bless_firth}
\end{figure}

Figure \ref{fig: ukbb_bless_firth} compares the raw age effect size images of our method BLESS, estimated via (a) approximate posterior sampling and (b) variational inference, to (c) the mass-univariate approach Firth regression. It should be noted that we omit the comparison to the other baseline method BSGLMM as the computation of the Bayesian spatial model becomes infeasible due to the large sample size of this study. We highlight how BLESS sufficiently regularizes the negligible age coefficients to zero while leaving the larger effects unaffected. This is a direct consequence of the structured spike-and-slab prior placed on the spatially-varying coefficients. Furthermore, the spatial MCAR prior allows the sparsity dictating parameters within the spike-and-slab prior to borrow strength from their respective neighboring voxels. We further illustrate this behavior by plotting the coefficients of the feature age of the entire 3D effect map of the brain in the scatterplots in Figure \ref{fig: ukbb_bless_firth}. The comparison between BLESS-VI and BB-BLESS coefficients again showcases the alignment of posterior mean estimates between the two parameter estimation procedures. The other scatterplots on the other hand capture the induced shrinkage of small effects to zero via BB-BLESS and BLESS-VI while the Firth regression parameter estimates vary for the negligible effects and exhibit non-zero values.

For inference, we threshold the test statistics of BB-BLESS  at a significance level of~5\%. In contrast, for BLESS-VI we threshold the posterior probability of inclusion at 0.5 in order to acquire its respective binary significance map. Hence, we exploit variable selection as a means to conduct inference. On the other hand, the mass-univariate approach Firth regression ignores any form of spatial dependence and hence requires the application of a multiple testing correction where we adjust the p-values with a FDR correction \citep{benjamini1995} at a significance level of 5\%. The results in Figure \ref{fig: ukbb_bless_firth} indicate a slightly larger extent of spatial activation for BB-BLESS and BLESS-VI compared to Firth regression for a sample size of $N=2,000$. For the covariate age, in the regression model estimated via Firth regression 10,171 voxels are deemed active based on uncorrected p-values. On the other hand, only 6,278 voxels pass the FDR adjusted threshold whereas in BB-BLESS 8,385 effect locations are detected by utilizing the full posterior to derive test statistics and similarly in BLESS-VI 8,257 effects are detected via simply thresholding the posterior inclusion probabilities. 

\subsection{Cluster Size Imaging Statistics}
\label{sec: cluster_size_imaging_statistics}
Cluster-extent based thresholding is the most commonly used inference technique for statistical maps in neuroimaging studies. By proposing BB-BLESS and hence by sampling from an approximate posterior, we are able to provide novel cluster size based imaging statistics, such as cluster size credible intervals, in addition to reliable uncertainty quantification of the spatially-varying coefficients. 

\begin{figure}[!ht]
\hspace{1.5cm}\rotatebox[origin=l]{90}{\footnotesize{(1) Cluster Size Inference}}
\begin{subfigure}{0.25\textwidth}
\includegraphics[width=1\linewidth]{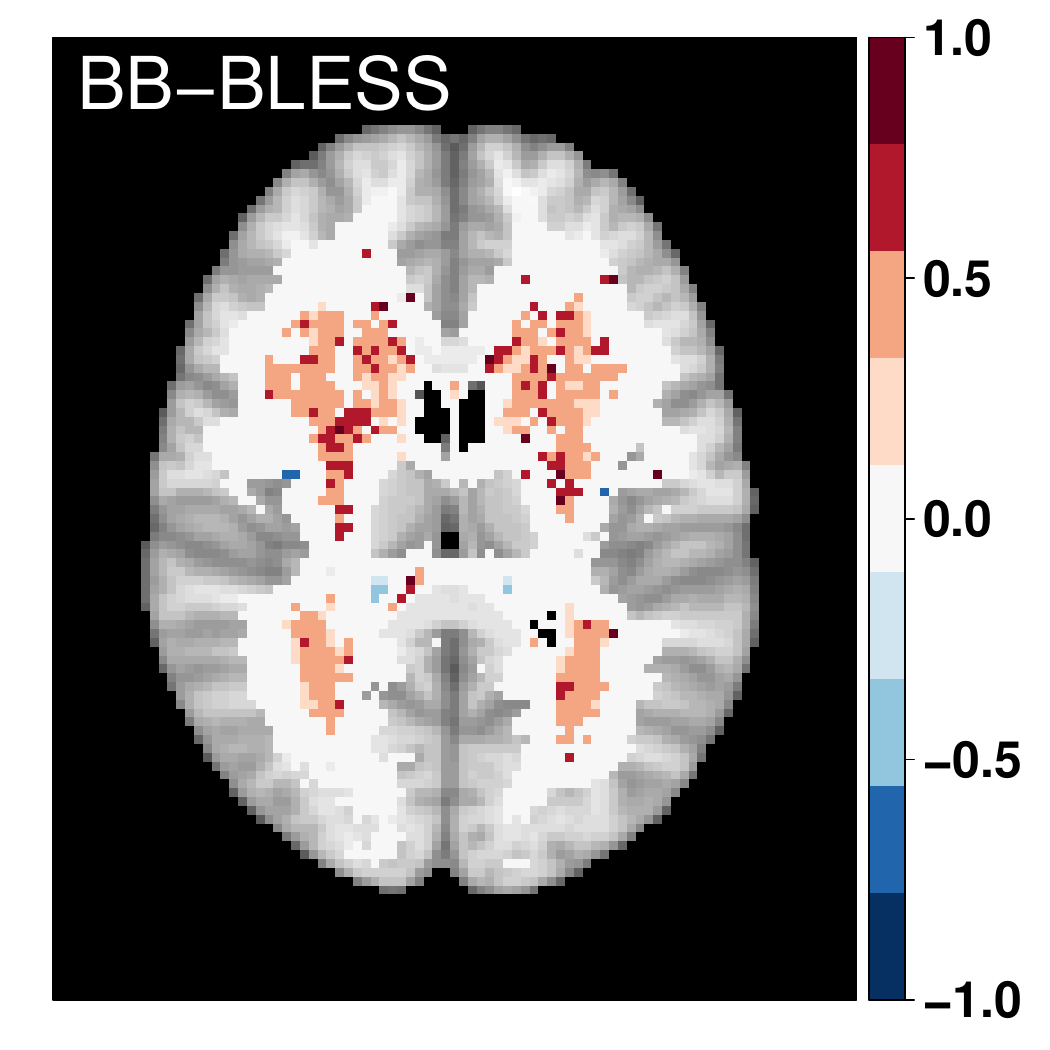}
\end{subfigure}
\begin{subfigure}{0.25\textwidth}
\includegraphics[width=1\linewidth]{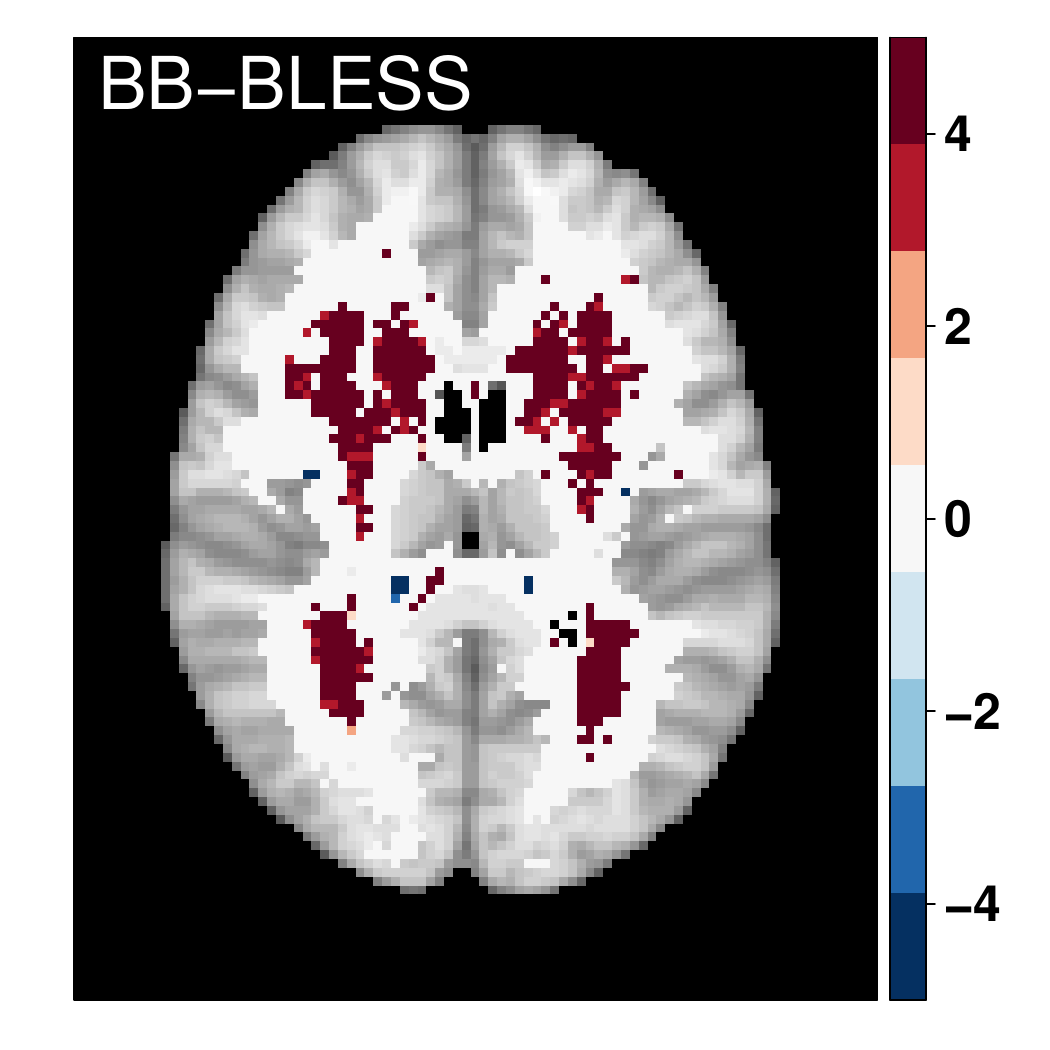}
\end{subfigure}
\begin{subfigure}{0.25\textwidth}
\includegraphics[width=1\linewidth]{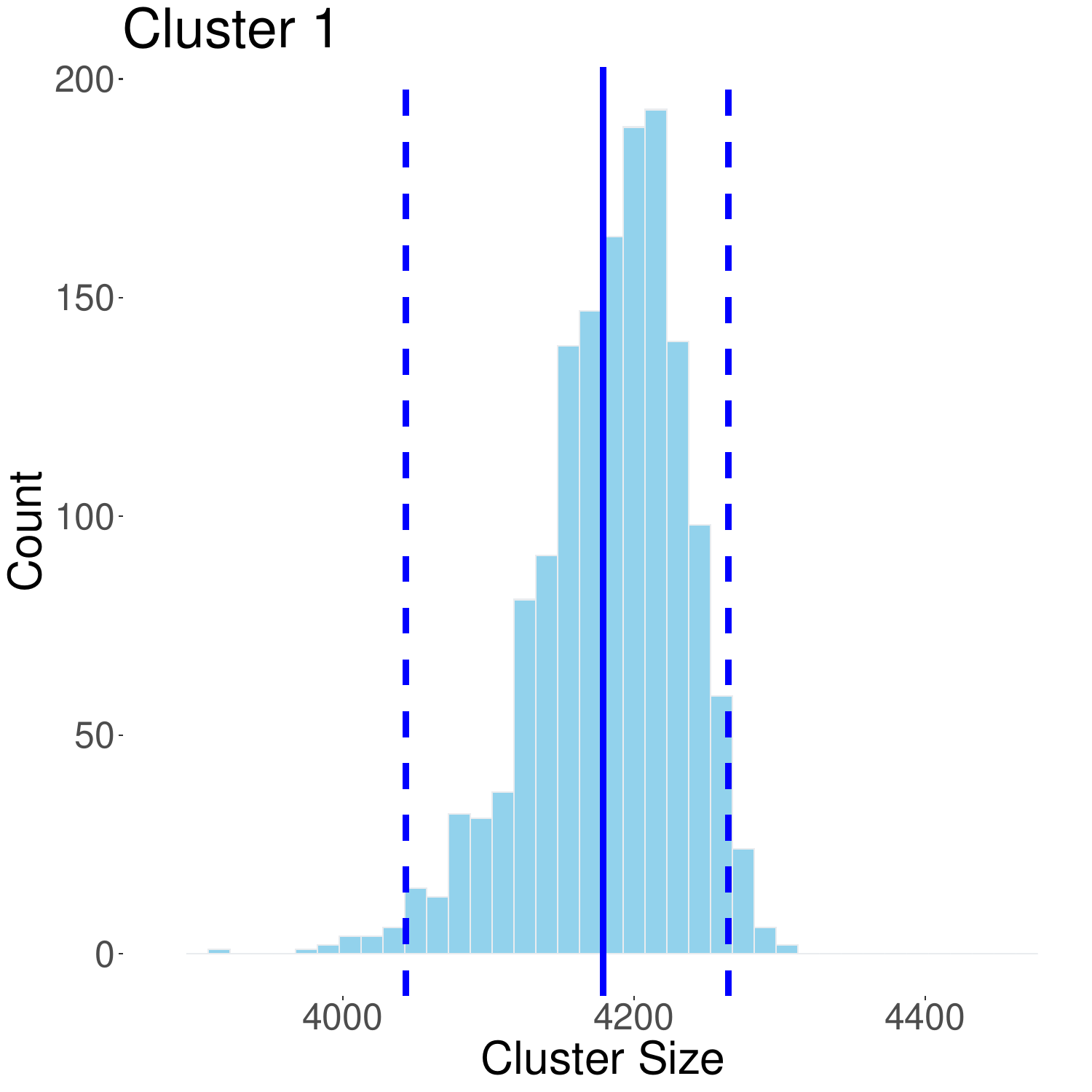}
\end{subfigure}

\hspace{1.5cm}\rotatebox[origin=l]{90}{\footnotesize{(2) Cluster Size Mapping}}
\begin{subfigure}{0.25\textwidth}
\includegraphics[width=1\linewidth]{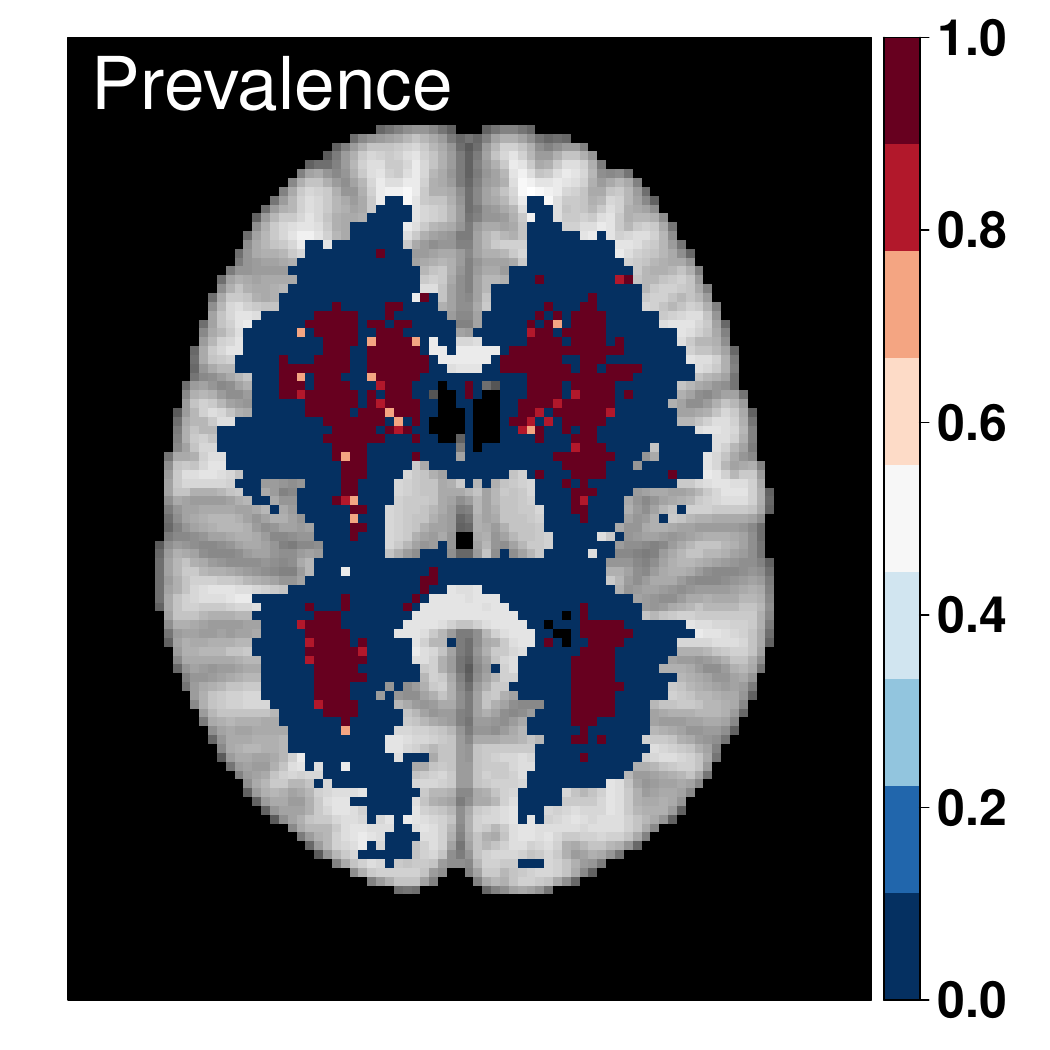}
\end{subfigure}
\begin{subfigure}{0.25\textwidth}
\includegraphics[width=1\linewidth]{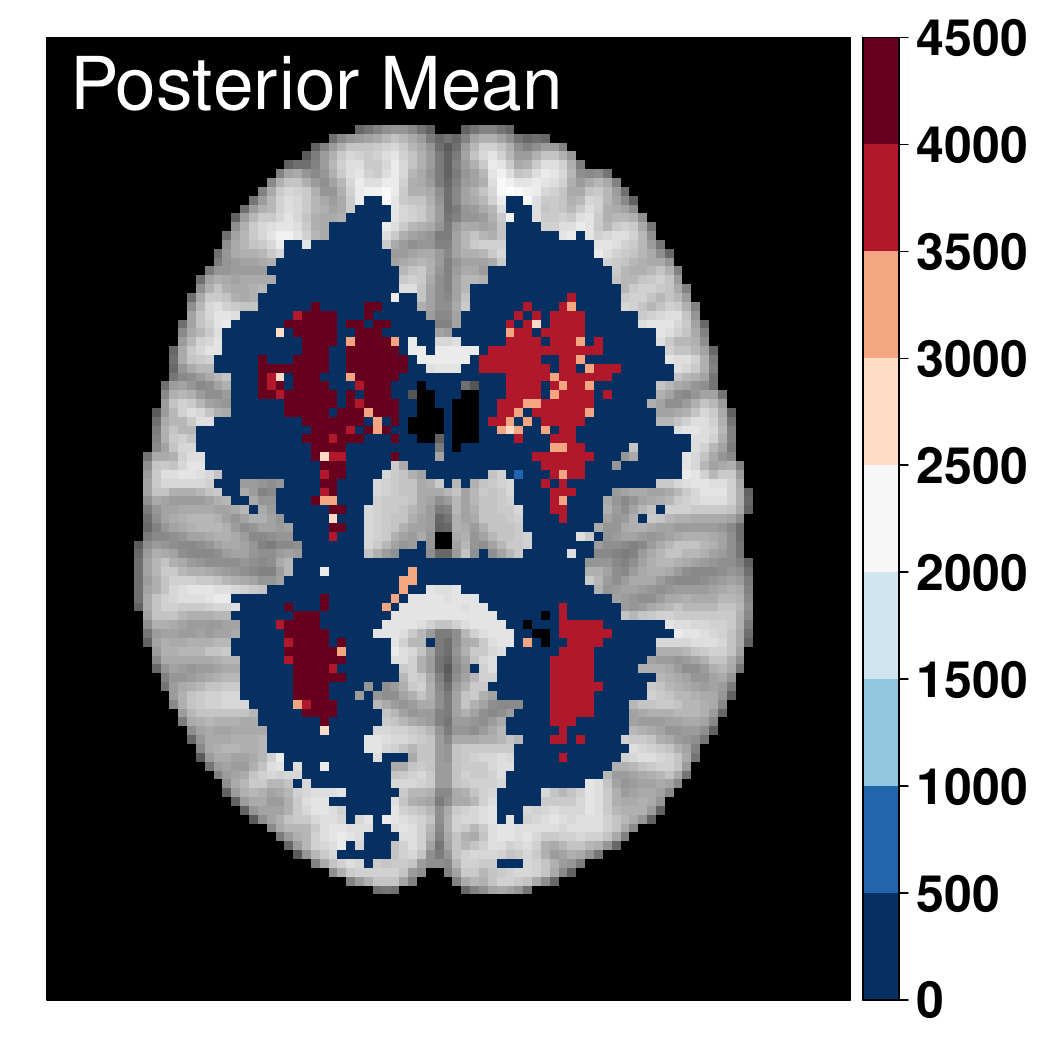}
\end{subfigure}
\begin{subfigure}{0.25\textwidth}
\includegraphics[width=1\linewidth]{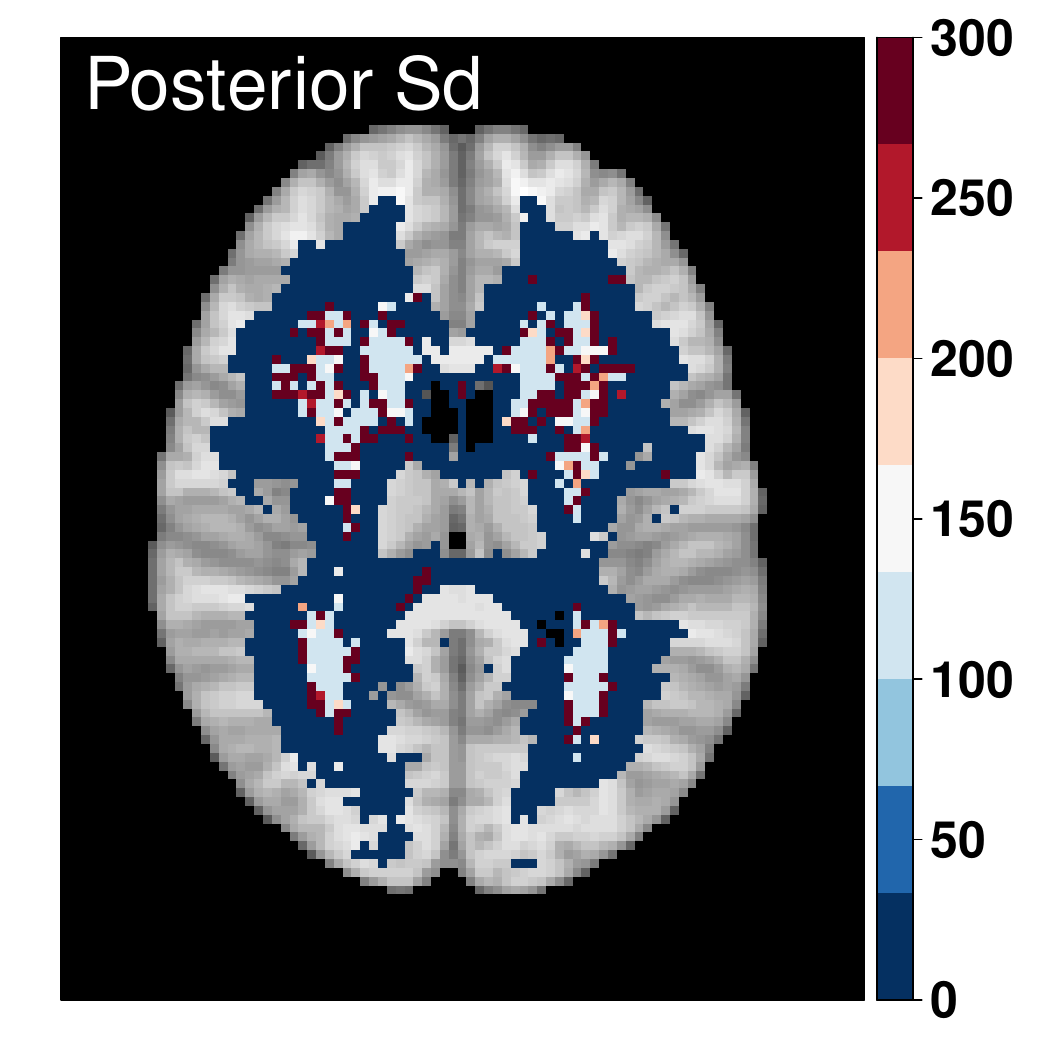}
\end{subfigure}
\caption{\textbf{(1) Cluster Size Inference:} Top left: Raw age effect size image. Top middle: Test statistic map for age effect. Top right: Cluster size distribution for the largest cluster detected by a cluster defining threshold of 2.3 (The solid line indicates the observed cluster size from BLESS-VI and the dashed lines signify the 95\% credible interval of cluster size.). \textbf{(2) Cluster Size Mapping:} Lower left, middle, right: Prevalence, posterior mean and posterior standard deviation map of cluster size, where the latter two statistics are determined for instances where the prevalence map exceeds a probability of 50\%. The prevalence map here indicates that both clusters have reliably large effects with values close to 1.}
\label{fig: cluster_approaches}
\end{figure}

We have shown that the raw age effect size image of the posterior mean for BB-BLESS in Figure \ref{fig: ukbb_bless_firth} obtained as an average over the bootstrap replicates $\Bar{\beta}(s_j) = \frac{1}{B} \sum_{b=1}^B \hat{\beta}^{(b)}(s_j)$ is almost identical to BLESS-VI. From our simulation studies, we expect the point estimates from both approximate methods to be identical. Moreover, BB-BLESS is able to capture the uncertainty of coefficients more reliably than BLESS-VI. This is important not only for providing uncertainty quantification at a population level but also for evaluating the predictive performance via posterior predictive checks and ensuring model robustness via calibration plots (see Section 4.4 of the supplementary material). With the latter we show that our model is well calibrated in predictive uncertainty and hence the model performance for BLESS does not change to a large extent when using new out-of-sample data. The approximate posterior samples can also be utilized to calculate cluster size based imaging statistics which require test statistics in their estimation. The statistical map in the top middle part of Figure \ref{fig: cluster_approaches}, acquired by the standardization of the raw effects $t(s_j)=\Bar{\beta}(s_j) / \sigma_{\beta}(s_j)$ with the posterior standard deviation $\sigma_{\beta}(s_j)$, shows that voxelwise inference based on thresholding the posterior probabilities of inclusion from BLESS-VI at~0.5 is similar to thresholding test statistics at a significance level of $\alpha=5\%$. 

We now highlight two novel cluster size approaches, based on cluster size inference and cluster size mapping, that can be calculated via BB-BLESS. In the first approach we acquire credible intervals of cluster size by utilizing the more accurate posterior standard deviation estimates of BB-BLESS to standardize the bootstrap samples. The resampled statistical maps are then thresholded by a cluster-defining threshold of 2.3 (equivalent to thresholding p-values at a significance level of 0.01) which generates cluster size maps for every bootstrap replicate. We then build a distribution of cluster size by identifying the intersection of each bootstrap cluster with the observed one and recording its respective cluster size. A distribution over cluster sizes for any cluster within the brain allows us to calculate an array of statistical quantities, such as credible intervals of cluster size. In the top right part of Figure \ref{fig: cluster_approaches}, we display the cluster size distribution for the largest cluster identified across the brain alongside its 95\% credible interval which ranges between a cluster size of 4,063 and 4,265 voxels and contains the observed cluster size value of 4,179 voxels. 

In a second cluster size mapping approach, we compute the voxelwise posterior probability of the standardized effect exceeding 2.3. This allows us to create a map of not just large effect but reliably large effect voxels. Comparing the cluster map with the occurrence map provides a measure for the reliability of cluster occurrence at a particular location within the brain. Due to the large spatial extent of the effect of age across the brain and viewing the central axial slice of the 3D maps, almost all voxel locations have a cluster prevalence close to 1. For these locations we then report posterior mean and standard deviation of reliable cluster size at locations where the prevalence of a cluster exceeds 50\%.

To summarize, Figure \ref{fig: ukbb_bless_firth} highlights how BLESS is able to reduce the identification of spurious associations for high-dimensional problems by shrinking the model's negligible coefficients to zero and leaving larger effects unaffected. In the UK Biobank, where we study how age is associated with occurrence of lesions, age has a very large effect on lesion incidence. However, many studies require methods to identify much subtler risk factors for lesion incidence and BLESS is able to identify these smaller effects alike with a higher level of specificity and sensitivity compared to the mass-univariate approach. Our methods aid a more accurate spatial localization of effects where we show that the effect of age on lesion incidence predominantly covers periventricular and deep white matter regions. Hence, our model provides us with a tool to identify the brain regions impacted by lesion occurrence within a large population and to determine the impact on cognitive, sensory, or autonomic loss potentially induced by a higher lesion burden due to increased age. In Figure 6 in the supplementary materials we also highlight the change of lesion incidence across the brain in 1000 out-of-sample subjects under 50 years and over 75 years old. While the lesion location is still focused around the ventricles, the predicted lesion incidence increases greatly with age which validates previous research findings \citep{KINDALOVA2021118090}. 

Moreover, we are also able to provide uncertainty estimates of parameter maps which help in the assessment of spatial associations at a population level, attaching risk assessments for identified biomarkers for diseases, and enabling the acquisition of cluster size based imaging statistics. The UK Biobank application for analyzing the effect of age on lesion occurrence only identifies two big clusters, which validates our expectation based around the magnitude of the effect for age. More importantly, BB-BLESS has the unique advantage to provide us with prevalence statements of cluster size quantities. A spatial map that can aid decisions for follow-up studies, when resources are scarce and a researcher needs to know the reliability of large effect voxels and cluster occurrence across the brain.
\section{Discussion}
\label{sec: discussion}
We have proposed a novel Bayesian spatial generalized linear model with a structured spike-and-slab prior for the analysis of binary lesion data. Our main contribution to the neuroimaging community is the development of a scalable version of a Bayesian spatial model that is able to diminish spurious associations by shrinking negligible coefficients to zero, to increase model interpretation via Bayesian variable selection and to provide a model that is also easily extendable to other neuroimaging modalities, such as functional MRI with a continuous response variable. 

The computational tractability of our method is also facilitated by using a data augmentation approach for the probit model and an analytical approximation to estimate the parameters in the logistic function in the Bernoulli prior on the latents within the spike-and-slab distribution \citep{albert1993, Jaakkola2000}. For future work, switching the probit link to a logit link enables the interpretation of the spatially-varying coefficients via log-odds ratios. Moreover, advances in Bayesian inference for efficient posterior estimation of logistic regressions using P\'{o}lya-Gamma latent variables circumvent the approximation by \cite{Jaakkola2000} and therefore have potential for gains in accuracy and computational efficiency \citep{polson2013, durante2019}.

Lastly, we would like to address that we limit our application of BLESS to a clinical application aiming at identifying the association between age and lesion incidence in a large scale population health study, the UK Biobank. However, it is a well established finding in the analysis of white matter hyperintensities that age is one of the strongest predictors of lesion incidence \citep{wardlaw2013}. Therefore, the study of more subtle risk factors for disease and the application of our cluster size based imaging statistics poses an interesting future research direction, as for example the further exploration of the cognitive impact of cerebrovascular risk-related white matter lesions \citep{veldsman2020}. 
\section*{Acknowledgments}
We thank Sahra Ghalebikesabi, Lorenzo Pacchiardi and Edwin Fong for~valuable~\mbox{feedback}. AM is funded by \mbox{EPSRC} StatML CDT (EP/S023151/1), \mbox{Oxford}-Radcliffe scholarship and Novartis, TEN by the Wellcome Trust (100309/Z/12/Z) and NIH grant 1R01DA048993, CH by The Alan Turing Institute, Health Data Research UK, the Medical Research Council UK, the EPSRC via the Bayes4Health grant EP/R018561/1, and AI for Science and Government UKRI and HG by Novartis. We would also like to thank the UK Biobank participants for their contribution to the study which was conducted under application 34077 and 8107.

\begingroup
\setstretch{1.0}
\footnotesize
\bibliography{manuscript}
\endgroup

\end{document}